\documentclass[prd,tightenlines,nofootinbib,superscriptaddress]{revtex4}

\usepackage{pre_all}
\usepackage{pre_CS}
\allowdisplaybreaks

\hypersetup{
pdfstartview = {FitH},
}
\hypersetup{
	colorlinks=true,         
	linkcolor=brown,          
	citecolor=red,        
	urlcolor=blue            
}

\begin{document}

\title{Complex Chern-Simons Theory with $k=8\N$ and An Improved Spinfoam Model with Cosmological Constant}

\author{{\bf Muxin Han}}\email{hanm@fau.edu}
\affiliation{Department of Physics, Florida Atlantic University, 777 Glades Road, Boca Raton, FL 33431, USA}
\affiliation{Institut f\"ur Quantengravitation, Universit\"at Erlangen-N\"urnberg, Staudtstr. 7/B2, 91058 Erlangen, Germany}

\author{{\bf Qiaoyin Pan}}\email{qpan@fau.edu}\thanks{corresponding author}
\affiliation{Department of Physics, Florida Atlantic University, 777 Glades Road, Boca Raton, FL 33431, USA}

\date{\today}

\begin{abstract}

This paper presents an improvement to the four-dimensional spinfoam model with cosmological constant ($\Lambda$-SF model) in loop quantum gravity. The original $\Lambda$-SF model, defined via $\SL(2,\mathbb{C})$ Chern-Simons theory on graph-complement 3-manifolds, produces finite amplitudes and reproduces curved 4-simplex geometries in the semi-classical limit. However, extending the model to general simplicial complexes necessitated ad hoc, non-universal phase factors in face amplitudes, complicating systematic constructions. We resolve this issue by redefining the vertex amplitude using a novel set of phase space coordinates that eliminate the extraneous phase factor, yielding a universally defined face amplitude. Key results include: (1) The vertex amplitude is rigorously shown to be well-defined for Chern-Simons levels $k \in 8\mathbb{N}$, compatible with semi-classical analysis ($k \to \infty$). (2) The symplectic structure of the Chern-Simons phase space is modified to accommodate $\SL(2,\mathbb{C})$ holonomies, relaxing quantization constraints to $\mathrm{Sp}(2r,\mathbb{Z}/4)$. (3) Edge amplitudes are simplified using constraints aligned with colored tensor models, enabling systematic gluing of 4-simplices into complexes dual to colored graphs. (4) Stationary phase analysis confirms consistency of critical points with prior work, recovering Regge geometries with curvature determined by $\Lambda$. These advancements streamline the spinfoam amplitude definition, facilitating future studies of colored group field theories and continuum limits of quantum gravity. The results establish a robust framework for 4D quantum gravity with non-zero $\Lambda$, free of previous ambiguities in face amplitudes.
\end{abstract}

\maketitle

\tableofcontents

\section{Introduction}

The introduction of a non-vanishing cosmological constant to the covariant loop quantum gravity (LQG), also called the spinfoam model \cite{Rovelli:2014ssa,Perez:2012wv}, is essential to define a finite theory with bounded spinfoam amplitudes. For instance, in three dimension (3D), the Turaev-Viro (TV) model \cite{Turaev:1992hq,Archer:1991rz} is defined based on the quasi-Hopf algebra $\UQ$ with $q=e^{2\pi i G_{\rm 3D}\hbar\sqrt{\Lambda}}$ (taking the speed of light $c=1$) being a root-of-unity that encodes a positive cosmological constant $\Lambda$, where $G_{\rm 3D}$ is the gravitational constant in 3D. It can be viewed as a regularized version of the Ponzano-Regge spinfoam model \cite{Ponzano:1968se}, deforming the symmetry algebra $\su(2)$ to  $\UQ$ with only finite-dimensional irreducible representations hence producing finite amplitudes. In In four dimension (4D), the spinfoam model with $\Lambda\neq 0$ introduced in \cite{Han:2021tzw}, denoted as $\Lambda$-SF model in this paper, also possesses the property of finiteness. It can be viewed as the regularized version of the Engle-Pereira-Rovelli-Livine (EPRL) model \cite{Engle:2007wy}. 

The $\Lambda$-SF model is defined based on the $\SL(2,\bC)$ Chern-Simons theory with a complex coupling constant $t=k+is$ on a graph-complement 3-manifold, where $k\in\Z_+$ and $s=\gamma k$ with $\gamma\in\R$ being the Barbero-Immirzi parameter. The Chern-Simons level $k$ encodes the value of $\Lambda$ by $k=\f{3}{2G_{\rm 4D}\hbar\gamma|\Lambda|}$ where $G_{\rm 4D}$ is the gravitational constant in 4D. More precisely, the vertex amplitude of the $\Lambda$-SF model is a constrained Chern-Simons partition function defined using quantum Teichm\"uller theory method \cite{EllegaardAndersen:2011vps,Dimofte:2011gm}. 
The feature that the vertex amplitude therein reproduces the Regge action of a constantly curved 4-simplex at the semi-classical regime is another evidence that this spinfoam model is promising for studying 4D quantum gravity with a non-zero $\Lambda$. 

Recent developments of this spinfoam model include generalizing the definition of spinfoam amplitude from a 4-simplex to simplicial complexes, which requires proper definitions of edge amplitudes associated to the internal curved tetrahedra and face amplitudes associated to the internal curved triangles of a 4-complex. However, in the attempts of \cite{Han:2023hbe,Han:2024reo} that take the $\Lambda$-SF vertex amplitude as the building block, a non-trivial phase factor has to be added to the face amplitude in addition to a quantum dimension as the quantum deformation of the EPRL face amplitude, so that the semi-classical approximation of the full amplitude has an unambiguous geometrical interpretation. Such a factor does not show up in the face amplitude of EPRL model. What is worse is that the phase factor can not be universally defined but takes different forms for different 4-complexes and has to be computed case by case. This brings difficulty in defining the spinfoam amplitude for a general 4-complex in a systematical way. 

This cumbersome situation is rooted in the definition of the vertex amplitude of the $\Lambda$-SF model.  In order to get a well-behaved wave function for any Chern-Simons level $k\in \Z_+$, a particular set of conjugate momenta are chosen, which are then quantized to derivative operators on the Chern-Simons wave functions. The price to pay is that these momenta do not have a simple geometrical interpretation. Especially, the conjugate momentum to a position variable encoding a triangle area turns out to be the linear combination of the variable encoding the dihedral angle hinged by the triangle and other phase space variables. The added phase factor in the face amplitude is there to cancel these other variables so that the critical deficit angles hinged by internal triangles of the spacetime triangulation vanish, giving rise to the desired constantly curved geometry. We refer readers to \cite{Han:2024reo} for more details. 

In this paper, we improve the $\Lambda$-SF model by redefining the vertex amplitude by using a new set of phase space coordinates. It allows us to eliminate the phase factor in the face amplitude, making it universally defined, which will bring great convenience to the calculation of semi-classical spinfoam amplitude. (This new set of coordinates was also used in \cite{Han:2015gma}.) 
We prove that the Chern-Simons wave function, hence the vertex amplitude, can be well-defined for $k\in 8\N$. 
When the semi-classical regime is concerned, where $k\rightarrow \infty$ is considered, this restriction is not harmful. 

The restriction to $k\in8\N$ is relevant to lifting the gauge group $\PSL(2,\bC)$, on which the Chern-Simons partition function on the triangulation building block is defined as the solution to $\PSL(2,\bC)$ flat connection condition, to $\SL(2,\bC)$. 
It is reflected in the classical theory that the symplectic transformations are relaxed from being within ${\rm Sp}(2r,\Z)$ to ${\rm Sp}(2r,\f{\Z}{4})$ where $2r$ is the dimension of the Chern-Simons phase space. 
This feature does not show up in the example of Chern-Simons theory on knot-complement 3-manifolds (\eg \cite{Dimofte:2011ju,Dimofte:2011gm}), but it is necessary for the amplitude construction in our case studying Chern-Simons theory on a graph-complement 3-manifold. As a feature of it, the new momenta variables are coordinates of $\SL(2,\bC)$ moduli space of flat connection, even though the position variables can be restricted to the $\PSL(2,\bC)$ counterparts. Geometrically, these $\SL(2,\bC)$ variables are closely related to the boosts between local 3D frames of different tetrahedra in a 4-simplex. 

Compared to \cite{Han:2024reo}, we also simplify the definition of the edge amplitude which is similar to the one used in \cite{Han:2023hbe}. Such edge amplitude restricts the way of gluing 4-simplices in composing a 4-complex. It is valid for 4-complexes whose dual graph is one on which {\it colored tensor models} are defined \cite{Gurau:2011aq,Gurau:2011xp}, also called the {\it colored graph}. 
The $1/N$ expansion ($N$ being the size of the tensor) of the colored tensor model is dominated by graphs dual to the triangulation of spheres. This spinfoam amplitude could be used to study the {\it colored group field theory} (CGFT) \cite{Gurau:2011aq} of spinfoam, which can expectedly be expressed as a (weighted) summation of spinfoam amplitudes for all spacetime triangulations with colored spinfoam graphs. 
CGFT could be an approach to regain the diffeomorphism symmetry at the continuum limit, potentially with some restriction on the spacetime topology. Nevertheless, the simplification of the edge amplitude and face amplitude would bring benefits for this purpose. 
Using the stationary phase analysis, we show that the new definition of the spinfoam amplitudes leads to the same critical points as in \cite{Han:2023hbe,Han:2024reo} for a general 4-complex whose spinfoam graph is consistent with a colored graph. 

This paper is organized as follows. In Section \ref{sec:vertex_amplitude}, we construct the vertex amplitudes, in a different way from \cite{Han:2021tzw,Han:2023hbe,Han:2024reo}, of the spinfoam model, starting from the classical setting and re-defining the Chern-Simons partition function on the $\SG$ 3-manifold. We then construct the edge amplitudes and the face amplitudes for colored graphs in Section \ref{sec:colored_graph_complex}. We analyze the critical points of the spinfoam amplitude in Section \ref{sec:critical_point}. We conclude and discuss possible applications of this framework in Section \ref{sec:conclusion}.

\section{Vertex amplitude for spinfoam model with $\Lambda\neq 0$}
\label{sec:vertex_amplitude}

\begin{figure}[h!]
\centering
\includegraphics[width=0.3\textwidth]{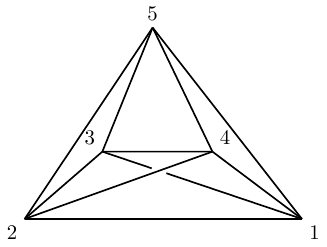}
\caption{$\Gamma_5$ graph projected on $\R^2$. }
\label{fig:Gamma5}
\end{figure}

We start from building the vertex amplitude, which is the amplitude for a constantly curved 4-simplex. The geometry of a constantly curved 4-simplex can be captured by moduli space $\cM_\Flat(\partial(\SG),\SL(2,\bC))$ of flat connection on $\SG$ \cite{Haggard:2014xoa}, which is the $\Gamma_5$ graph complement of the 3-sphere $S^3$. The $\Gamma_5$ graph projected on $\R^2$ is illustrated in fig.\ref{fig:Gamma5}.
$\cM_\Flat(\partial(\SG),\SL(2,\bC))$ is the Chern-Simons solution space on $\SG$, which inspires us that the spinfoam amplitude can be defined based on the Chern-Simons partition function on $\SG$. In this paper, we construct the spinfoam amplitude in a similar way as in the $\Lambda$-SF model in terms of the symplectic coordinates of the moduli space $\cM_\Flat(\partial(\SG),\SL(2,\bC))$ of $\SL(2,\bC)$ flat connection, using techniques introduced in \cite{Dimofte:2011gm,Dimofte:2013iv,Dimofte:2014zga}. 
The difference mainly lies in that we choose a different set of symplectic coordinates to express the Chern-Simons partition function, which are better suited for generalization. To avoid tautology, we only sketch the steps identical to $\Lambda$-SF model, for which we refer readers to the original paper \cite{Han:2021tzw} and review in \cite{Han:2023hbe}, and stress the modified parts. 

We start from the classical setting in Section \ref{subsec:ideal_triangulation}, where we describe the Chern-Simons phase space $\cM_\Flat(\partial(\SG),\SL(2,\bC))$ with a chosen set of symplectic coordinates. We describe the quantization and construct the well-defined wave functions in the Hilbert space in Section \ref{subsec:quantization}. This allows us to step-by-step construct the Chern-Simons partition function on $\SG$ through Weil transformation of wave functions, which we illustrate in Section \ref{subsec:CS_partition}. Finally, in Section \ref{subsec:simplicity}, we impose the simplicity constraints on this partition function and obtain the vertex amplitude. 

\subsection{Ideal triangulation and flat connection}
\label{subsec:ideal_triangulation}

The moduli space $\cM_\Flat(\partial(\SG),\SL(2,\bC))$ of $\SL(2,\bC)$ flat connection is a lift of $\cM_\Flat(\partial(\SG),\PSL(2,\bC))$ on the boundary of $\SG$. Before the lift, the coordinates can be used to express $\PSL(2,\bC)\cong \SL(2,\bC)/\Z_2$ holonomies on $\SG$ based on a so-called {\it ideal triangulation} \cite{Fock:2003alg,Gaiotto:2009hg}. 
In this subsection, we first review the symplectic structures of $\cM_\Flat(\partial(\SG),\PSL(2,\bC))$ and $\cM_\Flat(\SG,\PSL(2,\bC))$. We remind the reader that the symplectic coordinates that we choose in this paper are different from those in the $\Lambda$-SF model. 

\begin{figure}[h!]
\centering
\begin{minipage}{0.45\textwidth}
\includegraphics[width=0.5\textwidth]{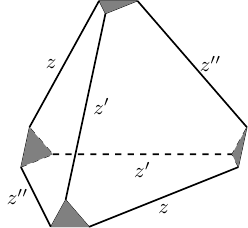}
\subcaption{}
\label{fig:ideal_tetra}
\end{minipage}
\begin{minipage}{0.45\textwidth}
\includegraphics[width=0.6\textwidth]{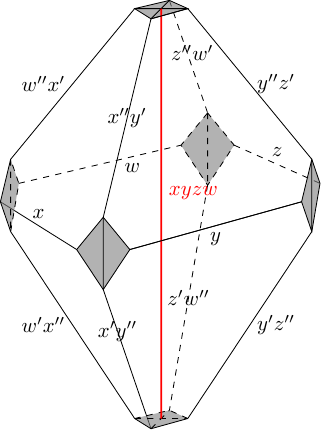}
\subcaption{}
\label{fig:ideal_octa}
\end{minipage}
\caption{{\it (a)} An ideal tetrahedron. {\it (b)} An ideal octahedron composed by 4 ideal tetrahedra. Cusp boundaries are shaded in gray. }
\label{fig:ideal_oct_tetra}
\end{figure}

The ideal triangulation $\TSG$ of $\SG$ contains the triangulation of the {\it geodesic boundaries} created by removing the open balls around nodes of $\Gamma_5$ and the triangulation of the {\it cusp boundaries} created by removing the open tubular neighbourhood of links of $\Gamma_5$. We refer to \cite{Han:2021tzw,Han:2023hbe} for a more detailed description but only emphasize that $\TSG$ can be decomposed into 5 {\it ideal octahedra} as shown in fig.\ref{fig:ideal_octa}, each of which can be further decomposed into 4 {\it ideal tetrahedra} by adding an internal edge. 
An ideal tetrahedron, denoted as $\Delta$, is a tetrahedron whose vertices are truncated as shown in fig.\ref{fig:ideal_tetra}. It is the building block of $\TSG$ and the phase space $\cP_{\partial\Delta}=\cM_\Flat(\partial\Delta,\PSL(2,\bC))$ of Chern-Simons theory on $\partial\Delta$. 
Associate the flat connection $A$ in $\cP_{\partial\Delta}$ with a {\it framing flag}, which is a choice of flat section $s=\lb s^0,s^1 \rb^\top\in\bC^2$ in an associated $\bC\bP^1$ bundle over every cusp boundary satisfying $\rd s=As$. 
It is the Lagrangian submanifold of $\cP_{\partial \triangle}$ with {\it framed} flat connection, which can be formulated in terms of the so-called {\it Fock-Goncharov (FG) coordinates} $z,z',z''\in\bC^*$ associated to the edges (on the geodesic boundary) of $\Delta$ with a relative position as shown in fig.\ref{fig:ideal_tetra} by the following relation
\be
\cP_{\partial\Delta}=\{z,z',z''\in\bC^*|zz'z''=-1\}\in\lb \bC^*\rb^2\,.
\label{eq:P_ideal_tetra_bdy}
\ee
The constraint $zz'z''=-1$ on the FG coordinates allows us to describe $\cP_{\partial\Delta}$ only in terms of $z$ and $z''$ whose Poisson bracket (written in terms of their logarithms) reads\footnote{
One can also lift the coordinates $z,z'',\zt,\zt''$ of $\cP_{\partial\Delta}$ to $\sqrt{-z},\sqrt{-z''},\sqrt{-\zt},\sqrt{-\zt''}$, which are FG coordinates of $\cM_\Flat(\partial\Delta,\SL(2,\bC))$. We will discuss this set of coordinates in the next subsection.
}
\be
\{Z,Z''\}=1\,,\quad 
Z:=\ln z\,,\quad 
Z'':=\ln z''\,.
\label{eq:Poisson_ZZpp}
\ee
On top of that, $\cL_\Delta$ is an algebraic curve on $\cP_{\partial\Delta}$ defined as
\be
\cL_\Delta=\{(z,z''; \zt,\zt'')\in\cP_{\partial \triangle} \mid z^{-1}+z''-1=0, \zt^{-1}+\zt''-1=0\}=\cM_\Flat(\Delta,\PSL(2,\bC))\,.
\label{eq:A-poly}
\ee

Holonomies of flat connection along paths on the boundary of a 3-manifold can be calculated in terms of the FG coordinates by the {\it snake rule} \cite{Gaiotto:2009hg,Dimofte:2013lba} (see also Appendix D of \cite{Han:2023hbe} for a concise description of the snake rule used in this spinfoam model). 
In this case, the $\SL(2,\bC)$ holonomy $h$ along a loop surrounding each cusp boundary of $\Delta$, gives
\be
h=\mat{cc}{1&0\\
zz'(z^{-1}+z''-1)&-zz'z''}\stackrel{\cL_\Delta}{=}\id_{\SL(2,\bC)} \,.
\ee
This reflects the fact that the pair of FG coordinates $(z,z'')$ does describe a flat connection on $\Delta$. 

On an ideal octahedron, four copies of FG coordinates $(x,x',x''),(y,y',y''),(z,z',z''),(w,w',w'')$ are associated to the edges as in fig.\ref{fig:ideal_octa}, each of which is assigned to an ideal tetrahedron submanifold in the same way as in fig.\ref{fig:ideal_tetra}. When an edge $E$ is shared by different $\Delta$'s, the FG coordinates $\{x_E^\Delta\}_\Delta$ from different $\Delta$'s are multiplied without ordering. For an internal edge, the FG coordinates are subject to a constraint $\prod_{\Delta\ni E}x_E^\Delta=1$. Therefore,
\be
xyzw=1
\quad\Longleftrightarrow\quad
X+Y+Z+W=2\pi i
\ee
for the internal edge when the FG coordinates are assigned in the way of fig.\ref{fig:ideal_octa}. 
Here, $x=e^X,y=e^Y,z=e^z,w=e^W$. 
This constraint eliminates one set of the FG coordinates, say $(w,w',w'')$, for one $\Delta$, and the symplectic coordinates are provided by $(\fZ,P_{\fZ}:=\fZ''-W'')$, where $\ln \fz\equiv\fZ\in\{X,Y,Z\}$, with Poisson bracket $\{\fZ_1,P_{\fZ_2}\}=\delta_{\fZ_1,\fZ_2}$. We will use this notation extensively in this paper. 

The above constructions contribute to the holomorphic part of the complex Chern-Simons theory on which the spinfoam model is based. Similar results are obtained with the anti-holomorphic FG coordinates $\tilde{\fZ}:=\ln \tilde{\fz}$ and $\tilde{\fZ}'':=\ln\tilde{\fz}''$, which take the same Poisson bracket $\{\tilde{\fZ},\tilde{\fZ}''\}=1$ as in \eqref{eq:Poisson_ZZpp} and construct an algebraic curve on $\cP_{\partial\Delta}$ through $\tilde{\fz}^{-1}+\tilde{\fz}''-1=0$. On an ideal octahedron, four copies of anti-holomorphic FG coordinates $(\xt,\xt''),(\yt,\yt''),(\zt,\zt''),(\wt,\wt'')$ on four copies of $\cP_{\partial\Delta}$'s are subject to the same constraint $\xt\yt\zt\wt=1$ or equivalently $\Xt+\Yt+\Zt+\Wt=2\pi i$ when written in terms of their logarithmic counterparts. 

We will use an equivalent set of coordinates $(\mu,\nu)\in\bC^2$ and periodic $(m,n)\sim(m+k,n+k)$ to parametrize $\fZ,\,\fZ''$ and their anti-holomorphic counterparts $\tfZ,\,\tfZ''$ such that\footnotemark{}
\be
\fZ=\frac{2 \pi i}{k}(-i b \mu-m)\,, \quad 
\fZ''=\frac{2 \pi i}{k}(-i b \nu-n)\,,\quad 
\tfZ=\frac{2 \pi i}{k}\left(-i b^{-1} \mu+m\right)\,,\quad 
\tfZ''=\frac{2 \pi i}{k}\left(-i b^{-1} \nu+n\right)\,,
\label{eq:paramtrize_ZZpp}
\ee
where $b$ is a phase defined in terms of the Barbero-Immirzi parameter $\gamma$ by
\be
b^2=\frac{1-i \gamma}{1+i \gamma}, \quad \re(b)=:\f{Q}{2}>0, \quad \im(b) \neq 0, \quad|b|=1\,.
\label{eq:b}
\ee
In a special case when $(\mu,\nu)\in\R^2$, $\tfZ$ is the complex conjugate of $\fZ$. 
The imaginary parts of $\mu$ and $\nu$, denoted as $\alpha:=\im(\mu)$ and $\beta:=\im(\nu)$ are fixed and the pair $(\alpha,\beta)$ is called the positive angle of a $\triangle$. Positive angle is important for the quantum theory, especially for convergence of the wave function. We refer to \cite{Han:2021tzw,Dimofte:2014zga} for more details. 
\footnotetext{
The use of $4\pi i$ instead of $2\pi i$ in the expression \eqref{eq:paramtrize_ZZpp} reflects that $(\fz,\fz'',\tilde{\fz},\tilde{\fz}'')$ are coordinates on $\cM_\Flat(\partial\Delta,\PSL(2,\bC))\subset\cM_\Flat(\partial\Delta,\SL(2,\bC))$. This is different from the parametrization used in \cite{Han:2021tzw}, where no lift of moduli space was considered. Such parametrization is important for the quantum theory construction in the current paper. }

$\TSG$ is composed of 5 ideal octahedra with no extra internal edges. Therefore, the phase space coordinates of $\cP_{\partial(\SG)}=\cM_\Flat(\partial(\SG),\PSL(2,\bC))$ is a collection of 5 copies of phase space coordinates of ideal octahedra, denoted as 
\be
\vec{\Phi}=\lb X_i,Y_i,Z_i\rb^\top_{i=1,\cdots,5}\,,\quad 
\vec{\Pi}=\lb P_{X_i},P_{Y_i},P_{Z_i}\rb^\top_{i=1,\cdots,5}\,,
\quad \{\Phi_I,\Pi_J\}=\delta_{IJ}\,.
\label{eq:def_Phi_Pi}
\ee

In quantum theory, it is more convenient to work on symplectic coordinates localized to the 4-holed spheres or annuli of $\partial(\SG)$. It corresponds to a symplectic transformation of the original coordinates $(\vec{\Phi},\vec{\Pi})$. We choose such new symplectic coordinates that are localized on $\partial(\SG)$ as well as have simple geometrical interpretations. The new coordinates are determined by a symplectic matrix $\bM$ and vectors $\vec{t}_\alpha,\vec{t}_\beta$ by
\be
\mat{c}
{\vec{\cQ} \\\vec{\cP}}
=\bM\mat{c}{\vec{\Phi}\\\vec{\Pi}}+i\pi \mat{c}{\vec{t}_\alpha\\\vec{t}_\beta}\,,\,\,
\bM=\mat{cc}{\bA&\bB\\{\bf C}&{\bf D}}\,,
\label{eq:symp_transf}
\ee
where $\bA,\bB,{\bf C}$ and ${\bf D}$ matrices are all $15\times 15$ matrices with half-integer entries and $\vec{t}_\alpha$ and $\vec{t}_\beta$ are length-$15$ vectors with half-integer entries. Their explicit expressions are given in Appendix \ref{app:symplec_tranf}.

The components of $\vec{\cQ}$ can be separated into two types: $\{M_a\}_{a=1,\cdots,5}$, each associated to a 4-holed sphere $\cS_a$, and $\{2L_{ab}\}_{a<b}$, each associated to the annulus connecting 4-holed spheres $\cS_a$ and $\cS_b$. Their conjugate momenta, which are components of $\vec{\cP}$, are $\{P_a\}_{a=1,\cdots,5}$ and $\{T_{ab}\}_{a<b}$ respectively. $(M_a,P_a)$ are called the FG coordinates on $\cS_a$ and $(2L_{ab},T_{ab})$ are called the {\it Fenchel-Nielsen (FN) length} and {\it FN twist} on the annulus $(ab)$ respectively. Geometrically, $2L_{ab}$ is related to the area of the triangle dual to $(ab)$ and $T_{ab}$ is related to the dihedral angle between two tetrahedra hinged by the triangle. 

$2L_{ab}$ and $T_{ab}$ are both calculated using the ``snake rule'' \cite{Dimofte:2011gm} and the explicit expressions in terms of $\vec{\Phi}$ and $\vec{\Pi}$ are given in \eqref{eq:L} and \eqref{eq:T} (see also Appendix C, E and H of \cite{Han:2023hbe} for the summary of the construction and snake rule). It can be shown using the Poisson bracket in \eqref{eq:def_Phi_Pi} that $\{2L_{ab},T_{cd}\}=\delta_{(ab),(cd)}$. $M_a$ and $P_a$ are both linear combinations of the 6 FG coordinates $\chi^{(a)}_{ij}$'s, each associated to an edge of $\TcS{a}$ ($ij$ in the subscript denotes that the edge is shared by ideal octahedra $\Oct(i)$ and $\Oct(j)$). 
We demand that they satisfy the Poisson brackets
\be
\{M_a,P_b\}=\delta_{ab}\,,\quad
\{M_a,T_{bc}\}=\{P_a,T_{bc}\}=\{M_a,M_b\}
=\{M_a,L_{ab}\}=\{P_a,L_{ab}\}=\{P_a,P_b\}=0\,.
\label{eq:Possion_M_P}
\ee
We solve 
\be
\sum_{(ij)}c^{(a)}_{ij}\{\chi_{ij}^{(a)},T_{bc}\}=0\,,\quad \forall\,a=1,\cdots,5
\ee
for coefficients $c^{(a)}_{ij}$ and find two independent solutions for each $a$ denoted by $c^{(a),1}_{ij}$ and $c^{(a),2}_{ij}$. We then let 
\be
M_a=\sum_{(ij)} c_{ij}^{(a),1}\chi^{(a)}_{ij}\,,\quad
P_a=\sum_{(ij)} c_{ij}^{(a),2}\chi^{(a)}_{ij}\,.
\ee
Explicitly,

\be\begin{aligned}
M_1&=\frac{1}{2} \left(-P_{Y_2}+P_{Y_3}-P_{Y_5}-2 P_{Z_3}+P_{Z_4}+P_{Z_5}-Y_2+Y_3-Y_5-Z_2-Z_3+2 Z_4+Z_5+2 i \pi \right)\,,\\
M_2&=\frac{1}{2} \left(-P_{X_3}+P_{X_5}+P_{Y_1}+P_{Y_3}+P_{Y_4}+2 X_5+Y_1+Y_4+Y_5+Z_1+Z_4+Z_5\right)\,,\\
M_3&=\frac{1}{2} \left(P_{X_2}-P_{X_5}+P_{Y_4}-P_{Z_1}-P_{Z_4}-2 X_1-2 X_5-2 Y_1+Y_2+Y_4-Y_5-2 Z_1+Z_2-Z_4-Z_5+6 i \pi \right)\,,\\
M_4&=\frac{1}{2} \left(P_{X_2}-P_{X_5}+P_{Y_5}+P_{Z_1}-P_{Z_2}+P_{Z_3}-Y_2+2 Y_5+2 Z_1-Z_2+i \pi \right)\,,\\
M_5&=\frac{1}{2} \left(P_{X_1}-P_{X_2}-P_{X_4}+P_{Y_3}-P_{Z_1}+P_{Z_2}-P_{Z_3}-2 X_4-Y_1+Y_2+Y_3-Y_4-Z_1+Z_2-Z_3-Z_4+6 i \pi \right)\,,\\
P_1&=\frac{1}{2} \left(P_{Y_2}-P_{Y_3}-P_{Y_4}-P_{Z_2}+P_{Z_3}-P_{Z_5}+Y_2-Y_3-Y_4+Z_2+Z_3-Z_4-2 Z_5+3 i \pi \right)\,,\\
P_2&=\frac{1}{2} \left(-P_{X_1}+2 P_{X_3}-P_{X_4}-P_{X_5}-P_{Y_3}+P_{Y_5}-2 X_1+2 X_3-2 X_4-2 X_5-Y_1+Y_3-Y_4-Z_1+Z_3-Z_4+i \pi \right)\,,\\
P_3&=\frac{1}{2} \left(-P_{X_1}+P_{X_4}-P_{X_5}+P_{Y_2}-P_{Z_4}+P_{Z_5}-2 X_1-Y_1+Y_2-Y_4+Y_5-Z_1+Z_2-Z_4+Z_5+5 i \pi \right)\,,\\
P_4&=\frac{1}{2} \left(-P_{X_3}+P_{X_5}-P_{Y_1}-P_{Y_2}+P_{Z_2}-P_{Z_3}-P_{Z_5}-2 X_3-Y_1+Y_2-Y_3-Y_5-Z_1+Z_2-Z_3-Z_5+5 i \pi \right)\,,\\
P_5&=\frac{1}{2} \left(P_{X_1}-P_{X_3}-P_{X_4}-P_{Y_1}+2 P_{Y_3}+P_{Z_2}-P_{Z_3}+P_{Z_4}-2 Y_1+Y_3+Y_4-Z_3+Z_4+2 i \pi \right)\,.
\end{aligned}\ee

It can be checked that \eqref{eq:Possion_M_P} are all satisfied with such solutions. This set of solutions was also used in \cite{Han:2015gma}. Then the matrices $\bA,\bB,{\bf C},{\bf D}$ and vectors $\vec{t}_\alpha,\vec{t}_\beta$ can be computed. Their explicit expressions are given in \eqref{eq:AB} and \eqref{eq:t_alpha_t_beta}. 
As a symplectic matrix, $\bM$ can be decomposed into matrices of symplectic generators. Notice that $\bA$ is invertible, the decomposition is

\be
\mat{cc}{\bA & \bB \\ {\bf C} & {\bf D}}
=\mat{cc}{\id & 0 \\ {\bf C}\bA^{-1} & \id}
\mat{cc}{0 & -\id \\ \id & 0}
\mat{cc}{\id & 0 \\ -\bB\bA^\top & \id}
\mat{cc}{(\bA^{-1})^\top & 0 \\ 0 & \bA}
\mat{cc}{0 & \id \\ -\id & 0}\,.
\label{eq:symplectic_decompse}
\ee
 
Here ${\bf C}\bA^{-1}$ and $-\bB\bA^\top$ are both symmetric matrices. 
Each of the generators gives rise to a Weil transformation on the partition function which we will describe in Sec.\ref{subsec:CS_partition}.

\subsection{Quantization and Weil transformations of wave functions}
\label{subsec:quantization}

To build the quantum theory, we promote the phase space variables $(\vec{\Phi},\vec{\Pi})$ to operators such that the parameters of $\vec{\Phi}$ are quantized to multiplication operators while parameters of $\vec{\Pi}$ are quantized to derivative operators or shift operators. 
For the reason that will be clear later, we consider the case $k=8N$ with $N\in\Z_+$ throughout this paper. We also lift $\cM_\Flat(\partial(\SG),\PSL(2,\bC))$ to $\cM_\Flat(\partial(\SG),\SL(2,\bC))$ and consider one conjugate pair of coordinates therein, which are parametrized as
\be
z=e^Z=\exp\left[\f{2\pi i}{k}\lb-ib\mu-m \rb \right]\,,\quad
z''=e^{Z''}=\exp\left[\f{2\pi i}{k}\lb-ib\nu-n \rb\right]\,,
\ee
whose parameters $\mu,\nu\in\bC$ and $m\sim k,n\sim k$ are periodic. $\mu,\nu$ are quantized to operators $\bmu,\bnu$ and $e^{\f{2\pi i}{k}m}, e^{\f{2\pi i}{k}n}$ are quantized to operators $e^{\f{2\pi i}{k}\bfm},e^{\f{2\pi i}{k}\bfn}$ satisfying 
\be
[\bmu,\bnu]=\f{k}{2\pi i}\,,\quad 
e^{\f{2\pi i}{k}\bfn}e^{\f{2\pi i}{k}\bfm}=e^{\f{2\pi i}{k}}e^{\f{2\pi i}{k}\bfm}e^{\f{2\pi i}{k}\bfn}\,,\quad
[\bmu,e^{\f{2\pi i}{k}\bfn}]=[\bnu,e^{\f{2\pi i}{k}\bfm}]=0\,.
\label{eq:comm_mu_nu}
\ee
The spectra $\mu,\nu$ of $\bmu,\bnu$ are complex\footnote{Technically speaking, only the real parts of $\mu,\nu$ are quantized while their imaginary parts are kept classical. See the description below \eqref{eq:b}.} while the spectra of $\bfm,\bfn$ respectively are discrete and bounded, \ie $l,r\in \Z/k\Z$.
Together with the tilde sectors
\be
\tilde{z}=e^{\widetilde{Z}}=\exp\left[\f{2\pi i}{k}\lb-ib^{-1}\mu+m \rb \right]\,,\quad
\tilde{z}''=e^{\widetilde{Z}''}=\exp\left[\f{2\pi i}{k}\lb-ib^{-1}\nu+n\rb\right]\,,
\ee
they are equivalent to the $q$-Weyl algebra (\resp $\qt$-Weyl algebra) satisfied by the quantization $\zb,\zb''$ (\resp $\zbt,\zbt''$) of $z,z''$ (\resp $\tilde{z},\tilde{z}''$) respectively:
\be
\zb''\zb=q\zb\zb''\,,\quad
\zbt''\zbt=\qt\zbt\zbt''\,,\quad
\zbt''\zb=\zb\zbt''\,,\quad
\zb''\zbt=\zbt\zb''\,,\quad 
\ee
where $q=e^{h}:=e^{\f{4\pi i}{t}}$ and $\qt\equiv e^{\tilde{h}}:=e^{\f{4\pi i}{\bar{t}}}$ being the quantum parameters that encode the complex coupling constants $t,\bar{t}\in\bC$ of the Chern-Simons theory. It is represented in the Hilbert space 
\be
\cH_{k}=L^2(\R)\otimes \bC^{k}\,.
\ee
The operators act on any state $f(\mu|m)\in\cH_{k}$ by
\be
\zb f(\mu|m)=zf(\mu|m)\,,\quad
\zb''f(\mu|m)=f(\mu+ib|m-1)\,,\quad
\zbt f(\mu|m)=\zt f(\mu|m)\,,\quad
\zbt'' f(\mu|m) =f(\mu+ib^{-1}|m+1)\,.
\ee

Any state in $\cH_{k}$ admits the cyclic periodicity $f(\mu|m+k)=f(\mu|m)$. 
Equivalently, when writing the state $f(z,\tilde{z})$ as a function of $z,\tilde{z}$, the operators $\zb,\zb'',\zbt,\zbt''$ act on $f(z,\tilde{z})$ as 
\be
\zb f(z,\tilde{z})= zf(z,\tilde{z})\,,\quad
\zb'' f(z,\tilde{z})=f(q z,\tilde{z})\,,\quad
\zbt f(z,\tilde{z})=\tilde{z} f(z,\tilde{z})\,,\quad
\zbt'' f(z,\tilde{z})=f(z,\qt\tilde{z})\,.
\label{eq:zzpp_on_f_2}\ee

In the construction of the Chern-Simons partition function on $\SG$, we will also consider a sub-Hilbert space of $\cH_k$ as the invariant subspace of the operator algebra formed by $\zb,(\zb'')^2,\zbt,(\zbt'')^2$. This algebra has a non-trivial center containing  $e^{\pi i\bfm}$ as the following vanishing commutators hold.
\be\ba{c}
[e^{\pi i\bfm},\zb]=[e^{\pi i\bfm},(\zb'')^2]=[e^{\pi i\bfm},\zbt]=[e^{\pi i\bfm},(\zbt'')^2]=0\,.
\ea
\ee
As $ \lb e^{\pi i\bfm}\rb^2 f(\mu|m)\equiv f(\mu|m)$ by definition, the eigenvalues of  $e^{\pi i\bfm}$ are $\pm 1$. This means one can decompose $\cH$ into 2 invariant subspaces of these center operators:
\be
\cH\simeq \cH_{+}\oplus \cH_{-}\,,\quad
\cH_{\pm}\simeq L^2(\R)\otimes \bC^{4N}\,.
\ee
where each subspace $\cH_{\pm}$ is an irreducible representation of the operator algebra formed by $\zb,(\zb'')^2,\zbt,(\zbt'')^2$.
The decomposition is such that any state $f_{\pm}(\mu|m)\in\cH_{\pm}$ is the eigenstate of  $e^{\pi i\bfm}$ with eigenvalue $\pm$, \ie
\be
e^{\pi i\bfm}f_{\pm}(\mu|m)=\pm f_{\pm}(\mu|m)\,.
\ee
This means the range of $m$ is restricted in a non-trivial function $f_{\pm}(\mu|m)$. Specifically,
 \be
 f_{+}(\mu|m)\equiv e^{\pi i\bfm}f_{+}(\mu|m)=\begin{cases}
 f_{+}(\mu|m)\,, & m \text{ even}\\
 -f_{+}(\mu|m)\,, & m \text{ odd}
 \end{cases}
 \ee
 which forces $f_{+}(\mu|m)=0$ when $m$ is odd. On the other hand,
 \be
 f_{-}(\mu|m)\equiv- e^{\pi i\bfm}f_{-}(\mu|m)=\begin{cases}
 -f_{-}(\mu|m)\,, & m \text{ even}\\
 f_{-}(\mu|m)\,, & m \text{ odd}
 \end{cases}
 \ee
 which forces $f_{-}(\mu|m)=0$ when $m$ is even. Therefore, $f_{+}(\mu|m)$ is supported on even $m$ while $f_{-}$ is defined on odd $m$. This motivates us to write the function $f_{\pm}$ in terms of $\mu'$ and $m'$ such that
 \be
 m=\begin{cases}
 	2m'\,, & \text{for }f_+\\
 	2m'+1\,, & \text{for }f_-
 \end{cases}\,,\quad
 \mu=2\mu'\,,\quad\forall\,\text{for }f_\pm\,.
 \ee
 Then
 \be
 z=\begin{cases}
 \exp\left[\f{2\pi i}{4N}\lb -ib\mu'-m' \rb\right]\,,\quad & \text{for }f_+\\
 e^{-\pi i/k}\exp\left[\f{2\pi i}{4N}\lb -ib\mu'-m' \rb\right]\,,\quad & \text{for }f_-
 \end{cases}\,,\quad
 \tilde{z} =\begin{cases}
 \exp\left[\f{2\pi i}{4N}\lb -ib^{-1}\mu'+m' \rb\right]\,,\quad &\text{for }f_+\\
 e^{\pi i/k}\exp\left[\f{2\pi i}{4N}\lb -ib^{-1}\mu'+m' \rb\right]\,,\quad & \text{for }f_-
 \end{cases}\,.
 \label{eq:z_zt_parity}
 \ee
 The unitary map between $\cH_{+}$ and $\cH_{-}$ is given by $e^{\frac{2\pi i}{k}\bfn}$ which transforms $m\rightarrow m+1$.
 
The partition function for an ideal tetrahedron is given by solving the quantum constraints 
\be
\zb''+\zb^{-1}-1=0\,,\quad
\zbt''+\zbt^{-1}-1=0\,,
\label{eq:quantum_A_poly}
\ee
which are the quantization of the defining relation \eqref{eq:A-poly} for $\cL_\Delta$ and its tilde sector. 
As we have chosen $k\in\Z_+$ for simplicity, $\im(b)<0$ or equivalently $|q|<1$.
The solution to \eqref{eq:quantum_A_poly} gives the quantum dilogarithm function 
\be
\Psi_\triangle(\mu|m)=
	 \prod_{j=0}^\infty\f{ 1-\qt^{j+1}\zt^{-1}}{1-q^{-j}z^{-1}}\,.
\label{eq:quantum_dilog}
\ee
It forms the building block of the partition function on $\SG$.
The Chern-Simons partition function on $\SG$ is written as a function of coordinates $(\vec{\cQ},\vec{\cP})$, which is a symplectic transformation from coordinates $(\vec{\Phi},\vec{\Pi})$ on five copies of ideal octahedra with $\vec{\Phi}=(X_a,Y_a,Z_a)^\top_{a=1,\cdots,5}$ and $\vec{\Pi}=(P_{X_a},P_{Y_a},P_{Z_a})^\top_{a=1,\cdots,5}$ whose Poisson brackets are \cite{Han:2021tzw}
\be
\{\Phi_I,\Pi_J\}=\delta_{IJ}\,,\quad \forall I,J=1,\cdots,15\,.
\ee
Denote the parametrization of the coordinates $(\vec{\cQ},\vec{\cP})$ in terms of 
$\vec{\mu},\vec{\nu},\vec{m},\vec{n}$ in the same way as $(\vec{\Phi},\vec{\Pi})$:
\be
\vec{\cQ}=\f{2\pi i}{k}\lb -ib\vec{\mu}-\vec{m} \rb\,, \quad
\vec{\cP}=\f{2\pi i}{k}\lb -ib\vec{\nu}-\vec{n} \rb \,.
\ee
Also denote the corresponding parameters of $2L_{ab}$ (\resp $T_{ab}$) as $\mu_{ab}$ and $m_{ab}$ (\resp $\nu_{ab}$ and $n_{ab}$) while the corresponding parameters of $M_a$ (\resp $P_a$) as $\mu_a$ and $m_a$ (\resp $\nu_a$ and $n_a$). 

Let us start with a general discussion on the symplectic transformation of interest.
The symplectic transformation for a set of $2r$ symplectic coordinates can be separated into different types of generators, namely U-type, T-type, S-type transformation, each of which corresponds to a $2r\times 2r$ symplectic matrix as follows.
\begin{subequations}
\begin{align}
\text{U-type: }& \,{\bf U}(\bA)=\mat{cc}{\bA & {\bf 0}\\{\bf 0} & \lb{\bf A}^{-1}\rb^\top}\,,\,\,\, \bA^{-1}\in {\rm GL}(r,\f{\Z}{2})\,, \\
\text{T-type: }& \, {\bf T}(\bB)=\mat{cc}{{\bf 1} & {\bf 0}\\{\bf B} & {\bf 1} } \,,\quad \bB=\bB^\top \in M(r,\f{\Z}{4})\,,\\
\text{S-type: }& \, {\bf S}=\mat{cc}{{\bf 0} & -{\bf 1} \\ {\bf 1} & {\bf 0}}\,,
\end{align}
\label{eq:symp_transf_all}
\end{subequations}
where $M(r,\f{\Z}{4})$ denotes the set of $r\times r$ matrices with entries belonging to $\Z/4$. 
On top of that, two kinds of affine translations $\sigma_{\alpha}$ and $\sigma_{\beta}$, each labelled by a length-$r$ vector $\vec{t}$ with half-integer elements, are also involved in the symplectic transformation.  

When $\bA^{-1},\bB\in M(r,\Z)$, the symplectic transformations \eqref{eq:symp_transf_all} correspond to the Weil transformations \cite{weil1964certains,Dimofte:2011gm} on a wave function
$f(\vec{\mu}|\vec{m})\in\cH_{s_1}\otimes \cH_{s_2}\otimes\cdots\otimes\cH_{s_r}\subset\cH^r= \bigotimes_{I=1}^{r}\cH_I\equiv \bigotimes_{I=1}^{r}\lb \bigoplus_{s=\pm} \cH^I_{s}\rb$ with $s_I\in\pm $ that takes the following form. 

\begin{subequations}
\begin{align}
\text{U-type: } &\quad 
\lb {\bf U}(\bA)\triangleright f \rb(\vec{\mu}|\vec{m})=\f{1}{\sqrt{\det(\bA)}}f(\bA^{-1}\cdot \vec{\mu}|\bA^{-1}\cdot \vec{m})\,,
\label{eq:U-type}\\
\text{T-type: }&\quad
\lb {\bf T}(\bB)\triangleright f \rb(\vec{\mu}|\vec{m})=(-1)^{\vec{m}\cdot\bB\cdot\vec{m}}e^{\f{i\pi}{k}\lb-\vec{\mu}\cdot\bB\cdot\vec{\mu}+\vec{m}\cdot\bB\cdot\vec{m}\rb}f(\vec{\mu}|\vec{m})\,,
\label{eq:T-type}\\
\text{S-type: }&\quad
\lb {\bf S}\triangleright f \rb(\vec{\mu}|\vec{m})= \f{1}{k^r}\sum_{\vec{n}\in(\Z/k\Z)^r}\int_{\cC^{\times r}}\rd^r\nu \, e^{\f{2\pi i}{k}\lb-\vec{\mu}\cdot\vec{\nu}+\vec{m}\cdot\vec{n} \rb}f(\vec{\nu}|\vec{n})\,,\\
\text{affine translation for position: }&\quad 
\lb \sigma_{\alpha}(\vec{t})\triangleright f\rb(\vec{\mu}|\vec{m}) = f(\vec{\mu}-i\f{Q}{2}\vec{t}|\vec{m})\,,
\label{eq:affine_a}\\
\text{affine translation for momentum: }&\quad 
\lb \sigma_{\beta}(\vec{t})\triangleright f\rb(\vec{\mu}|\vec{m}) = e^{\f{\pi Q}{k}\vec{\mu}\cdot \vec{t}} f(\vec{\mu}|\vec{m})\,,
\label{eq:affine_b}
\end{align}
\label{eq:unitary_transf_all}
\end{subequations}
where the integration contour $\cC^{\times r}:=\R^r + i\vec{\beta}$ is shifted from the real space to the complex space determined by the positive angle structure. 
Note that the affine translation for position \eqref{eq:affine_a} and that for momentum \eqref{eq:affine_b} do not commute even though their symplectic transformations do. They differ by a pure phase for imaginary $h$ and $\tilde{h}$, in which case the Weil transformations \eqref{eq:unitary_transf_all} should be understood as projective representations. 

The S-type transformation can also be made on only one pair of conjugate variables given by
\be
{\bf S}_i=\mat{cc}{{\bf 0} & \diag(0,\cdots,\stackrel{i\text{-th}}{-1},\cdots,0)\\
\diag(0,\cdots,\stackrel{i\text{-th}}{1},\cdots,0) & {\bf 0}}\,.
\ee
We call this the partial S-type transformation and it transforms the wave function as
\be
({\bf S}_i\triangleright f)(\vec{\mu}|\vec{m}) =\f{1}{k}\sum_{n\in\Z/k\Z}\int_\cC\rd\nu \,  e^{\f{2\pi i}{k}\lb -\mu_i\nu+m_in\rb}
f(\mu_1,\cdots,\stackrel{i\text{-th}}{\nu},\cdots,\mu_r|m_1,\cdots,\stackrel{i\text{-th}}{n},\cdots,m_r)\,.
\ee

{In this paper, we generalize the U-type and T-type transformations to the case with matrices $\bA^{-1}\in {\rm GL}(r,\f{\Z}{2})$ and $\bB\in M(r,\f{Z}{4})$ as shown in \eqref{eq:symp_transf_all}. We will use the same expression but restrict the domain of the function $f(\vec{\mu}|\vec{m})\in \cH^{\otimes r}$ so that these transformations act unambiguously, meaning that the resulting function is single-valued on $(\Z/k\Z)^r$. }

{We first consider the U-type transformation with ${\bf U}(\bA)\equiv {\bf U}$, $\bA^{-1}$ containing half-integer entries. Then $\lb {\bf U}\triangleright f\rb(\vec{\mu}|\vec{m})$ is guaranteed to be well-defined only when $\vec{m}\in 2(\Z/4N\Z)^r$, in which case it is single-valued on $2(\Z/4N\Z)^r$ hence also single-valued on $(\Z/k\Z)^r$:
\be
f(\bA^{-1}\cdot\vec{\mu}|\bA^{-1}\cdot\lb \vec{m}+k\vec{\delta}_L\rb )=f({\bf A}^{-1}\cdot\vec{\mu}|{\bf A}^{-1}\cdot\vec{m})\,,\quad 
\vec{\delta}_L:=\lb 0,\cdots,0,\stackrel{L}{1},0,\cdots,0 \rb^\top\,,\quad \forall\,L=1,\cdots, r\,.
\ee
${\bf U}$ can be densely defined on $\cH^{\otimes r}$ as a projection 
\be
{\bf U}:\cH^{\otimes r}\rightarrow\cH^{\otimes r}_+\,.
\ee
On the other hand, T-type transformation with ${\bf B}={\bf B}^\top$ and ${\bf B}_{ij}\in\Z/4$ can not be densely defined on $\cH^{\otimes r}$. 
However, when $k=8N$ and $\vec{m}=4\vec{l}\in 4(\Z/2N\Z)^r$, it defines a single-valued function on $(\mathbb{Z}/k\mathbb{Z})^r$. To see that, we only need to show that $(-1)^{\vec{m}\cdot{\bf B}\cdot\vec{m}}e^{\frac{i\pi}{k}\vec{m}\cdot{\bf B}\cdot\vec{m}}\equiv e^{\frac{i\pi}{k}\vec{m}\cdot{\bf B}\cdot\vec{m}}$ is single-valued on $(\mathbb{Z}/k\mathbb{Z})^r$. 
\be\begin{split}
&e^{\frac{i\pi}{k}\left(\vec{m}+k\vec{\delta}_{I}\right)\cdot{\bf B}\cdot\left(\vec{m}+k\vec{\delta}_{I}\right)}
=\exp\left[\frac{i\pi}{k}\left(\vec{m}\cdot{\bf B}\cdot\vec{m}+2k\vec{m}\cdot\vec{{\bf B}}_{I}+k^{2}{\bf B}_{II}\right)\right]
=e^{\frac{i\pi}{k}\vec{m}\cdot{\bf B}\cdot\vec{m}}
e^{i\pi\left(2\vec{l}\cdot4\vec{{\bf B}}_{I}+2N4{\bf B}_{II}\right)}
=e^{\frac{i\pi}{k}\vec{m}\cdot{\bf B}\cdot\vec{m}}
\end{split}\ee
where $\vec{{\bf B}}_{I}={\bf B}\cdot\vec{\delta}_{I}\,,{\bf B}_{II}=\vec{\delta}_{I}\cdot{\bf B}\cdot\vec{\delta}_{I}$.
}

The S-type transformation and affine translations are the same as in \cite{Dimofte:2014zga} and is proven to be valid Weil transformations hence they also define single-valued function on $(\Z/k\Z)^r$. 
{For the above reasons, we consider $k=8N$ with $N\in\Z_+$ and $\vec{m}\in 4(\Z/2N\Z)^r$ in the rest of this paper.}


The finiteness of a wave function $f(\vec{\mu}|\vec{m})$ on a 3-manifold $\cM$ is supported by a non-empty positive angle structure, denoted as $\mathfrak{P}$, which is an open subset of the space spanned by $(\vec{\alpha},\vec{\beta}):=(\im(\vec{\mu}),\im(\vec{\nu}))\subset \R^{2r}$ Given a positive angle structure $\mathfrak{P}$, we define the functional space
\be
\mathcal{F}_{\mathfrak{P}}=\left\{\text { holomorphic } f: \mathbb{C}^N \rightarrow \mathbb{C} \mid \forall(\vec{\alpha}, \vec{\beta}) \in \mathfrak{P}, e^{-\frac{2 \pi}{k} \vec{\beta} \cdot \vec{\mu}} f(\vec{\mu}+i \vec{\alpha}) \in \mathcal{S}\left(\mathbb{R}^N\right) \text { is Schwartz class }\right\}\,.
\ee
Combining a discrete representation part $\left(\mathbb{C}^k\right)^{\otimes r}$, we define
\be
\mathcal{F}_{\mathfrak{P}}^{(k)}=\mathcal{F}_{\mathfrak{P}} \otimes_\mathbb{C}\left(\mathbb{C}^k\right)^{\otimes r}\,.
\ee
Let us only mention that the Chern-Simons partition function converges absolutely as long as the 3-manifold admits a non-empty positive angle structure, which will prove the boundedness of the vertex amplitude we are about to define, and refer interested readers to \cite{EllegaardAndersen:2011vps,andersen2013new,Dimofte:2014zga} for details of the proof.

\subsection{Chern-Simons partition function on $\SG$}
\label{subsec:CS_partition}

To have a simpler geometrical interpretation of the coordinates entering the Chern-Simons partition function, we use a different definition for the Chern-Simons partition function from \cite{Han:2023hbe}. The new partition function corresponds to the decomposition \eqref{eq:symplectic_decompse} of the symplectic matrix ${\bf M}$. In the decomposition, the matrices on the right-hand side give rise to the S-type, U-type, T-type, S-type and T-type transformations on the partition function respectively from right to left. 

Recall that the final position coordinates $\{\cQ_I\}_{I=1}^{15}$ and momentum coordinates $\{\cP_I\}_{I=1}^{15}$ can be separated into two classes locally dressing different parts of the boundary $\partial(\SG)$. The first 10 elements of $\vec{\cQ}$ are the FN lengths $\{2L_{ab}\}_{a<b}$, each assigned to an annulus $(ab)$ connecting $\cS_a$ and $\cS_b$ and their conjugate momenta are the FN twists $\{T_{ab}\}_{a<b}$ contributing to the first 10 elements of $\vec{\cP}$ which also dress the annuli of $\partial(\SG)$. Each of the last 5 elements of $\vec{\cQ}$ is an FG coordinate, denoted as $M_a$, dressing one 4-holed sphere $\cS_a$. Its conjugate momentum, denoted as $P_{a}$, is also an FG coordinate dressing $\cS_a$. 

We start with the partition function for 5 ideal octahedra 
\be
\cZ_\times(\vec{\mu}|\vec{m})
\equiv\cZ_\times(\{\mu_{X_i},\mu_{Y_i},\mu_{Z_i}\}_{i=1}^5|\{m_{X_i},m_{Y_i},m_{Z_i}\}_{i=1}^5)=\prod_{i=1}^5\cZ_{\Oct(i)}(\mu_{X_i},\mu_{Y_i},\mu_{Z_i}|m_{X_i},m_{Y_i},m_{Z_i})
\,,
\label{eq:Z_x}
\ee
where
\be
\cZ_{\Oct(i)}=\Psi_\Delta(\mu_{X_i}|m_{X_i})\Psi_\Delta(\mu_{Y_i}|m_{Y_i})\Psi_\Delta(\mu_{Z_i}|m_{Z_i})\Psi_\Delta(iQ-\mu_{X_i}-\mu_{Y_i}-\mu_{Z_i}|-m_{X_i}-m_{Y_i}-m_{Z_i})\,.
\ee 
$Q=b+b^{-1}$ was defined in \eqref{eq:b}. 
Here, we adopt the parametrization 
\be
\vec{\Phi}=\frac{2\pi i}{k}\lb-ib\, \vec{\mu}_0-\vec{m}_0\rb\,,\quad
\vec{\Pi}=\frac{2\pi i}{k}\lb-ib\, \vec{\nu}_0-\vec{n}_0\rb\,,\quad
\ee
where 
\be\ba{ll}
\vec{\mu}\equiv
\{\mu_{X_i},\mu_{Y_i},\mu_{Z_i}\}_{i=1}^5
\in\bC^{15}\,,\quad
&\vec{\nu}\equiv
\{\nu_{P_{X_i}},\nu_{P_{Y_i}},\nu_{P_{Z_i}}\}_{i=1}^5
\in\bC^{15}\,,\quad \\[0.15cm]
\vec{m}\equiv\{m_{X_i},m_{Y_i},m_{Z_i}\}_{i=1}^5
\in (\Z/k\Z)^{15}\,,\quad
&\vec{n}
\equiv\{n_{P_{X_i}},n_{P_{Y_i}},n_{P_{Z_i}}\}_{i=1}^5
\in (\Z/k\Z)^{15}\,.
\ea\ee
We first perform the inverse S-type transformation on $\cH^{\otimes 15}$, which corresponds to the Fourier transform on the wave function:
\be
\cZ_1(\vec{\mu}|\vec{m})
 =\f{1}{k^{15}}\int_{\R^{15}+i\vec{\beta}_1}\rd^{15}\nu\sum_{\vec{n}\in \lb\Z/k\Z\rb^{15}}e^{\f{2\pi i}{k}\lb \vec{m}\cdot\vec{n}-\vec{\mu}\cdot\vec{\nu}\rb}\cZ_\times(\vec{\nu}|\vec{n})\,,
\label{eq:Z0_2_Z1}
\ee 
The positive angle structure $\fP_\times$ for $\cS_\times$ is known to be non-empty \cite{Han:2021tzw}. 
When $\lb\vec{\alpha}_0,\vec{\beta}_0\rb\in\fP_\times$, $\lb\vec{\alpha}_1,\vec{\beta}_1\rb=\lb\vec{\beta}_0,-\vec{\alpha}_0\rb\in\fP_1$ hence the positive angle is satisfied and $\cZ_1\in\cF_{\fP_1}^{(k)}$. 

Secondly, we perform the U-type transformation given by the matrix ${\bf U}(\lb\bA^{-1}\rb^\top)$.
As all entries of $\bA$ are half-integers, to have a well-defined U-type transformation and a T-type transformation that follows, we let
\be
\cZ_2(\vec{\mu}|\vec{m})=\begin{cases}
\sqrt{\det(\bA)}\cZ_1(\bA^\top\cdot \vec{\mu}|\bA^\top\cdot \vec{m})\,, & \vec{m}\in 4(\Z/2N\Z)^{15}\\
0\,, &\text{otherwise}
\end{cases}\,,
\label{eq:def_Z2}
\ee
where $\sqrt{\det(\bA)}=8i$. 
That is, we restrict to the case when $\vec{m}=4\vec{l}\in 4(\Z/2N\Z)^{15}$ from now on. 
When $\lb\vec{\alpha}_1,\vec{\beta}_1\rb\in\fP_1$, $\lb\vec{\alpha}_2,\vec{\beta}_2\rb=\lb\lb\bA^{-1}\rb^\top\vec{\alpha}_1,\lb\bA^{-1}\rb^\top\vec{\beta}_1\rb\in\fP_2$ hence the positive angle is satisfied and $\cZ_2\in\cF_{\fP_2}^{(k)}$. 

Thirdly, the T-type transformation given by matrix ${\bf T}(-\bB\bA^\top)$ gives
\be
\cZ_3(\vec{\mu}|\vec{m})= e^{\f{i\pi}{k}\lb \vec{\mu}\cdot\bB\bA^\top\cdot \vec{\mu}-\vec{m}\cdot\bB\bA^\top\cdot \vec{m}\rb} \cZ_2(\vec{\mu}|\vec{m})\,.
\ee
As all entries of the symmetric matrix ${\bf T}(-\bB\bA^\top)$ are $\Z/4$, $\cZ_3$ is a well-defined wave function in $\cH^{\otimes 15}$. 
When $\lb\vec{\alpha}_2,\vec{\beta}_2\rb\in\fP_2$, $\lb\vec{\alpha}_3,\vec{\beta}_3\rb=\lb\vec{\alpha}_2,\vec{\beta}_2-\bB\bA^\top\vec{\alpha}_2\rb\in\fP_3$ hence the positive angle is satisfied and $\cZ_3\in\cF_{\fP_3}^{(k)}$. 

An inverse S-type transformation follows,  which is a Fourier transformation for $\cZ_3\in\cH^{\otimes 15}$. To prepare for the second T-type transformation that follows, we again restrict to the case of $\vec{m}\in 4(\Z/2N\Z)^{15}$. That is
\be
\cZ_4(\vec{\mu}|\vec{m}) =
\begin{cases}
\f{1}{k^{15}}\int\limits_{\R^{15}+i\vec{\beta}_4}\rd^{15}\nu\sum\limits_{\vec{n}\in 4\lb\Z/2N\Z\rb^{15}}e^{\f{2\pi i}{k}\lb  \vec{\mu}\cdot\vec{\nu}-\vec{m}\cdot\vec{n}\rb}\cZ_3(\vec{\nu}|\vec{n})\,, \quad & \vec{m}\in 4(\Z/2N\Z)^{15}\\[0.15cm]
0\,,\quad &\text{otherwise}
\end{cases}\,.
\ee 
When $\lb\vec{\alpha}_3,\vec{\beta}_3\rb\in\fP_3$, $\lb\vec{\alpha}_4,\vec{\beta}_4\rb=\lb\vec{\beta}_3,-\vec{\alpha}_3\rb\in\fP_4$ hence the positive angle is satisfied and $\cZ_4\in\cF_{\fP_4}^{(k)}$. 

Another T-type transformation given by ${\bf T}({\bf C}\bA^{-1})$ results in 
\be
\cZ_5(\vec{\mu}|\vec{m})= e^{\f{i\pi}{k}\lb -\vec{\mu}\cdot{\bf C}\bA^{-1}\cdot \vec{\mu}+\vec{m}\cdot{\bf C}\bA^{-1}\cdot \vec{m}\rb} \cZ_4(\vec{\mu}|\vec{m})\,.
\ee
Again, all entries of ${\bf C}\bA^{-1}$ are $\Z/4$ hence the sign factor $(-1)^{\vec{m}\cdot{\bf C}\bA^{-1}\cdot \vec{m}}$ is not needed and $\cZ_5$ is well-defined. 
When $\lb\vec{\alpha}_4,\vec{\beta}_4\rb\in\fP_4$, $\lb\vec{\alpha}_5,\vec{\beta}_5\rb=\lb\vec{\alpha}_4,\vec{\beta}_4+{\bf C}\bA^{-1}\vec{\alpha}_4\rb\in\fP_5$ hence the positive angle is satisfied and $\cZ_5\in\cF_{\fP_5}^{(k)}$. 

Lastly, an affine translation acts on $\cZ_5$ and gives the Chern-Simons partition function on $\SG$:
\be
\cZ_{\SG}(\vec{\mu}|\vec{m})
=e^{\f{\pi Q}{k}\vec{\mu}\cdot \vec{t}_\beta}\cZ_5(\vec{\mu}-i\f{Q}{2}\vec{t}_\alpha|\vec{m})\,.
\ee
It shifts the positive angle to $\lb\vec{\alpha},\vec{\beta}\rb=\lb\vec{\alpha}_5+\f{Q}{2}\vec{t}_\alpha,\vec{\beta}_5+\f{Q}{2}\vec{t}_\beta\rb$, whose requirement constitute $\fP_{\SG}$ and $\cZ_{\SG}\in\cF_{\fP_{\SG}}^{(k)}$. 

Combining the above steps, the Chern-Simons partition function takes the expression that

\begin{multline}
\mathcal{Z}_{S^{3}\backslash\Gamma_{5}}(\vec{\mu}|\vec{m})=\frac{8i}{k^{30}}e^{\frac{\pi Q}{k}\vec{\mu}\cdot\vec{t}_{\beta}}e^{\frac{i\pi}{k}\left(\vec{m}\cdot{\bf CA^{-1}}\cdot\vec{m}-\left(\vec{\mu}-i\f{Q}{2}\vec{t}_{\alpha}\right)\cdot{\bf CA^{-1}}\cdot\left(\vec{\mu}-i\f{Q}{2}\vec{t}_{\alpha}\right)\right)}\:\\
\sum_{\vec{n}\in4\left(\mathbb{Z}/2N\mathbb{Z}\right)^{15}}
\int{\rm d}^{15}\nu\,
\sum_{\vec{m}_0\in\left(\mathbb{Z}/k\mathbb{Z}\right)^{15}}\int{\rm d}^{15}\mu_0\:e^{\frac{i\pi}{k}\left(\vec{\nu}\cdot{\bf BA^{\top}}\cdot\vec{\nu}-\vec{n}\cdot{\bf BA^{\top}}\cdot\vec{n}\right)}e^{\frac{2\pi i}{k}\left(\left({\bf A}\vec{\mu}_0-\vec{\mu}+i\f{Q}{2}\vec{t}_{\alpha}\right)\cdot\vec{\nu}+\left(\vec{m}-{\bf A}\vec{m}_0\right)\cdot\vec{n}\right)}\mathcal{Z}_{\times}(\vec{\mu}_0|\vec{m}_0)\,.
\label{eq:SG_partition_final}
\end{multline}

The positive angle structure of $\cZ_{\SG}$ can then be expressed in terms of that for $\cZ_\times$:
\be
\left(\vec{\alpha},\vec{\beta}\right)
=\left(-\left({\bf A}^{-1}\right)^{\top}\vec{\alpha}_{0}-{\bf B}\vec{\beta}_{0}+\f{Q}{2}\vec{t}_{\alpha},-{\bf D}\vec{\beta}_{0}-{\bf C}\vec{\alpha}_{0}+\f{Q}{2}\vec{t}_{\beta}\right)\,,
\ee
where we have used the property of the symplectic matrix ${\bf M}$: ${\bf D}\bA^\top-{\bf C}\bB^\top=\id$. 
As no constraints are imposed on the ideal octahedra when they are glued to form the ideal triangulation of $\SG$, the positive angle structure remains non-empty after the Weil transformations. 

Having defined the Chern-Simons partition function for $\SG$, we need to impose the first-class constraints by restricting the FN lengths $\{L_{ab}\}_{a<b}$ and impose the second-class constraints by coupling with coherent states, one to each 4-holed sphere as in \cite{Han:2021tzw} to define a vertex amplitude. We will do this for one $\SG$ in the next subsection and generalize it to graph complement 3-manifolds $\cM\backslash\Gamma$ corresponding to more general 4-complexes all at once after defining the Chern-Simons partition function for $\cM\backslash\Gamma$ in the next section.

\subsection{Simplicity constraints and the vertex amplitude}
\label{subsec:simplicity}
To describe 4D quantum gravity out of the Chern-Simons theory, one should impose the {\it simplicity constraints} on the partition function \eqref{eq:SG_partition_final}. The simplicity constraints can be separated into two types: first-class constraints, which we impose strongly and second-class constraints, which we impose weakly so that non-trivial quantum states can exist as in the EPRL spinfoam model. The first-class constraints are imposed on $\cZ_{\SG}$ to satisfy \cite{Han:2021tzw}
\be
\re(\bmu_{ab})\cZ_{S^3\backslash\Gamma_5}(\vec{\mu}|\vec{m})=0\,,
\label{eq:1st_class_quantum}
\ee
where $\bmu_{ab}=\re(\bmu_{ab})+i\alpha_{ab}$ is the quantization of $\mu_{ab}$ (whose imaginary part is kept classical). This restricts $\re(\mu_{ab})=0$ in the argument of the partition function. Define the ``spin'' $j_{ab}$ such that
\be
2j_{ab}= m_{ab}\quad\rightarrow\quad
j_{ab}=0,2,\cdots,4N-2\,.
\label{eq:m_to_j}
\ee
$j_{ab}$ encodes the area $\fa_{ab}$ of the triangle $\triangle_{ab}$ by \cite{Haggard:2015ima,Han:2024reo} 
\be
\f{|\Lambda|}{3}\fa_{ab}=\begin{cases}
\f{4\pi }{k}j_{ab}\,,\quad & \text{ if } \,j_{ab}\in[0,2N-2]\\
\lb 2\pi-\f{4\pi }{k}j_{ab}\rb \,,\quad & \text{ if } \,j_{ab}\in[2N,4N-2]
\end{cases}\,. 
\label{eq:spin_to_area}
\ee
The quantum states satisfying the constraint \eqref{eq:1st_class_quantum} are then labelled by
\be
\cZ_{S^3\backslash\Gamma_5}(\{i\alpha_{ab}\}_{a<b}, \{\mu_{a}\}\mid \{j_{ab}\}_{a<b}, \{m_{a}\})\,.
\label{eq:Z_after_1st_constraints}
\ee 

The second-class constraints impose restrictions on the FG coordinates $\{M_a,P_a\}_a$ on 4-holed spheres, which are parametrized as
\be
M_a:=\f{2\pi i}{k}\lb -ib\mu_a-m_a \rb\,,\quad
P_a:=\f{2\pi i}{k}\lb -ib\nu_a-n_a \rb\,.
\ee
They are imposed weakly by coupling 5 coherent states, each associated to one $\cS_a$. Such a coherent state, denoted as $\Psi_{\hrho_a}(\re(\mu_a)|m_a)$, lives in the Hilbert space $\cH_{\cS_a}:=L^2(\R)\otimes \bC^k$ and is labeled by $\hrho_a:=(\zh_a,\xh_a,\yh_a)\in\bC\times[0,2\pi)\times [0,2\pi)$. Each label $\hrho_a$ corresponds to a point in the shape phase space of a tetrahedron with fixed triangle areas. 
The coherent state is defined as
\be
\Psi_{\hrho_a}(\re(\mu_a)\mid m_a):=\psi^0_{\zh_a}(\re(\mu_a))\otimes \xi_{(\xh_a,\yh_a)}(m_a)
\label{eq:coherent_state}
\ee
where

\begin{subequations}
\begin{align}
\psi_{\zh_a}^0(\mu)&:= \lb\f{2}{k}\rb^{1/4} e^{-\f{\pi}{k}\lb\mu-\f{k}{\pi\sqrt{2}}\re(\zh_a)\rb^2}e^{-i\sqrt{2}\mu\im(\zh_a)}\in L^2(\R)\,,
\label{eq:coherent_state_1}\\
\xi_{(\xh_a,\yh_a)}(m)&:=\lb\f{2}{k}\rb^{1/4} e^{\f{ik\xh_a\yh_a}{4\pi}}\sum_{p_a\in \Z}
e^{-\f{k}{4\pi}\lb \f{2\pi m}{k}-2\pi p_a-\xh_a \rb^2} e^{\f{ik}{2\pi}\yh_a\lb \f{2\pi m}{k}-2\pi p_a-\xh_a\rb}\in \bC^k\,.
\label{eq:coherent_state_2}
\end{align}\end{subequations}

It is easy to see that the coherent state peaks at (with chosen branches for $m_a$ and $n_a$, in which case $p_a=0$)
\be
\re(\mu_{a})=\f{k}{\sqrt{2}\pi}\re(\zh_{a})\,,\quad
m_{a}=\f{k}{2\pi}\xh_{a}\,,\quad
\re(\nu_{a})=-\f{k}{\sqrt{2}\pi}\im(\zh_{a})\,,\quad
n_{a}=-\f{k}{2\pi}\yh_{a}\,.
\label{eq:eom_position}
\ee
The simplicity constraints are imposed weakly in the sense that, in the semi-classical regime where the critical points \eqref{eq:eom_position} are reached, the corresponding coordinates $\vec{\cQ},\vec{\cP}$ are symplectic coordinates of $\cL_{\SG}=\cM_\Flat(\SG,\SU(2))$
, while the quantum theory is ``fuzzy" around the classical solution. 
The vertex amplitude is defined as

\be\begin{split}
\cA_v(\iota)&:=\la\prod\limits_{a=1}^5\bar{\Psi}_{\hrho_a}|\cZ_{S^3\backslash\Gamma_5}\ra\\
&=\sum_{\{m_a\}\in 4(\Z/2N\Z)^5}\int_{\R^5+i\beta_a} \rd^5\mu_a \,
\cZ_{S^3\backslash\Gamma_5}(\{i\alpha_{ab}\}_{a<b}, \{\mu_{a}+i\alpha_a\}\mid\{j_{ab}\}_{a<b}, \{m_{a}\})
\prod\limits_{a=1}^5\Psi_{\hrho_a}(\mu_a|m_a)\,,
\end{split}
\label{eq:vertex_amplitude}
\ee

where $\iota=(\{\alpha_{ab},j_{ab}\}_{a<b}, \{\hrho_a\}_{a=1}^5, \{\alpha_a,\beta_a\}_{a=1}^5)$, and we have specified the imaginary part $\alpha_a$ of the continuous parameter of $M_a$ hence $\mu_a\in\R$ in the expression. 

As we will describe in Section \ref{subsec:critical_point_vertex}, the vertex amplitude \eqref{eq:vertex_amplitude} peaks at the configuration $(\vec{\cQ},\vec{\cP})\in\cL_{\SG}$ that describes the geometry of a 4-simplex given consistent boundary condition. This means the geometry of a curved 4-simplex can be read from data of $\cL_{\SG}$, used to reconstruct the critical point of the vertex amplitude. This is explicitly explored in \cite{Pan:2025sut}.

\section{Spinfoam amplitude for colored graph}
\label{sec:colored_graph_complex}

After defining the spinfoam amplitude for one single 4-simplex, we would like to push this definition to more general (ideally all) 4-complex as triangulations of 4-manifolds. This can be done by further defining amplitudes for edges and faces of spinfoam graphs, called the edge amplitudes and face amplitudes respectively, that geometrically encode gluing of 4-simplices. 
It is related to the {\it group field theory} (GFT) \cite{Oriti:2006se,Oriti:2005mb,Oriti:2011jm,Rivasseau:2011hm} of spinfoam, which can be viewed as summing over amplitudes for all spacetime triangulations with a given boundary condition. GFT is a way to regain the diffeomorphism symmetry broken by considering a fixed triangulation in spinfoam.

A proposal for edge and face amplitudes was recently given in \cite{Han:2024reo}. They, however, do not take a very simple form (as in the EPRL model), and the face amplitudes can not be defined in a universal way. We will see that the new definition of the vertex amplitude in the previous section can eliminate the latter concern. Here, we consider the possibility of having a simple form of edge amplitude by considering a restricted way of gluing 4-simplices by dressing the edges of the spinfoam graph with ``colors''. The resulting GFT expressed as the sum of amplitudes that only includes such subset of triangulations is called the CGFT \cite{Gurau:2009tw}. It can be considered as a group generalization of the colored tensor model \cite{Gurau:2011aq,Gurau:2011xp,Gurau:2013cbh}\footnotemark{}, defined on the so-called colored graphs. 

Remarkably, amplitudes for colorable spinfoam graphs (which means the edges can be dressed to become colored graphs in the colored tensor model) can be defined in an exceptionally simple way, bringing benefits for computing amplitudes for complicated 4-complexes.
For this reason, we restrict ourselves to considering only the spinfoam amplitudes for spacetime triangulation dual to the (4+1)-colored graph. 
\footnotetext{
A 5-colored (independent identically distributed) tensor model is defined by the partition function \cite{Gurau:2011xp}
\be
\cZ_{N, \lambda, \bar{\lambda}}=\prod_{i=0}^4\left[\int\rd\phi^i_{n_i} \rd \bar{\phi}^i_{n_i}\right] e^{-S(\{\phi^i_{n_i} , \bar{\phi}^i_{n_i} \}_i)}\,,
\nn\ee
where $\phi^i_{n_i}$ is a rank-5 tensor with $i$ being the color, $\bar{\phi}^i_{n_i}$ its complex conjugate, $n_i$ is an abbreviation of the form $(n_{i\,i-1},\cdots,n_{i0},n_{i4},\cdots,n_{i\,i+1})$ where $n_{ij}\in\{1,\cdots, N\}$ for a given size $N\in\Z_+$, and $\lambda,\bar{\lambda}$ are the coupling constants. The measure $\rd\phi^i_{n_i} \rd \bar{\phi}^i_{n_i}$ is Gaussian-normalized and the action reads
\be
S(\{\phi^i_{n_i} , \bar{\phi}^i_{n_i} \}_i)=\sum_{i=0}^4 \sum_{n_i} \phi_{n_i}^i \bar{\phi}_{n_i}^i+\frac{\lambda}{N^{3}} \sum_n \prod_{i=0}^4 \phi_{n_i}^i+\frac{\bar{\lambda}}{N^{3}} \sum_n \prod_{i=0}^4\bar{\phi}_{n_i}^i\,.
\nn\ee
The summation $\sum_n$ in the interaction term is subject to the constraint $n_{ij}=n_{ji}$, which gives the restriction on the gluing for colored graph. 
}

\begin{figure}[h!]
\centering
\includegraphics[width=0.2\textwidth]{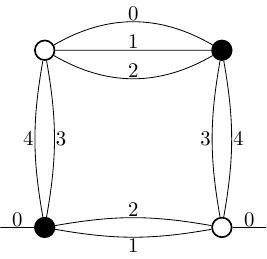}
\caption{An example of 5-colored graph.}
\label{fig:colored_graph}
\end{figure}
A 5-colored graph is a graph composed of 5-valent nodes such that each node is colored black or white and each of its incident links is colored by a distinct number $i\in\{0,1,\cdots,4\}$. When a link connects two nodes, the two nodes must have different colors. An example of a 5-colored graph is given in fig.\ref{fig:colored_graph}. In the spinfoam context, such graphs are spinfoam graphs dressed with colors. 
Each node is dual to a 4-simplex, each link is dual to a tetrahedron and each face, bounded by links with two different colors, is dual to a triangle. 
One can also view the duality in terms of the graph-complement manifold. Then each node is dual to an $\SG$, each link is dual to a 4-holed sphere and each face is dual to an annulus. 
We identify the label of a 4-holed sphere $\cS_a$ on $\partial(\SG)$ with the color of the edge in the colored graph by $a=i+1$. This gives a unique way to identify the 4-holed spheres of $\SG$'s when they are glued to form the graph-complement manifold. 

We first consider the gluing of $\cS_a$ and $\cS'_a$ from two different $\SG$'s, denoted by $\cS_a\sim\cS'_a$. {With no loss of generality, we let $\cS_a$ be from the $\SG$ corresponding to a white vertex of the colorable spinfoam graph and $\cS'_a$ from one corresponding to a black vertex therein. 
The gluing of a pair of 4-holed spheres is realized by first flipping the orientation of 4-holed sphere from $\SG$ of a black spinfoam vertex, $\cS'_a$ in this case, then identifying the six edges of the ideal triangulations, denoted by ${\bf T}(\cS_a)$ and ${\bf T}(\cS'_a)$, of the 4-holed spheres $\cS_a$ and $\cS'_a$ respectively, which leads to 6 independent gluing constraints. }
The flipping of the orientation also induces a sign change of Poisson brackets. Especially, $\{M'_a,P'_a\}=-1$. To unify the Poisson structure of the moduli space coordinates on $\cS_a$ and $\cS'_a$, we change the sign of the momentum $P'_a\rightarrow-P'_a$. 

The color of the spinfoam vertices can also be viewed as assigning the orientation to the spinfoam edges from white vertices to black vertices. This is key to maintain the same form of the vertex amplitudes for all spinfoam vertices while describing the gluing of 4-holed spheres in the Chern-Simons theory setting unambiguously. Another way is to change the definition of the vertex amplitude for black spinfoam vertices. In particular, one adds minus sign to some elements of $\vec{\cQ}$ and $\vec{\cP}$ hence changing the explicit expressions of the symplectic matrix ${\bf M}$ and vectors $\vec{t}_\alpha,\vec{t}_\beta$ in \eqref{eq:symp_transf_all}. This attempt was used for the melonic spinfoam graph in \cite{Han:2023hbe}. In this paper, our approach of defining the vertex amplitudes universally is closer to the construction in \cite{Han:2024reo}.

\begin{figure}[h!]
	\centering
\includegraphics[width=0.48\textwidth]{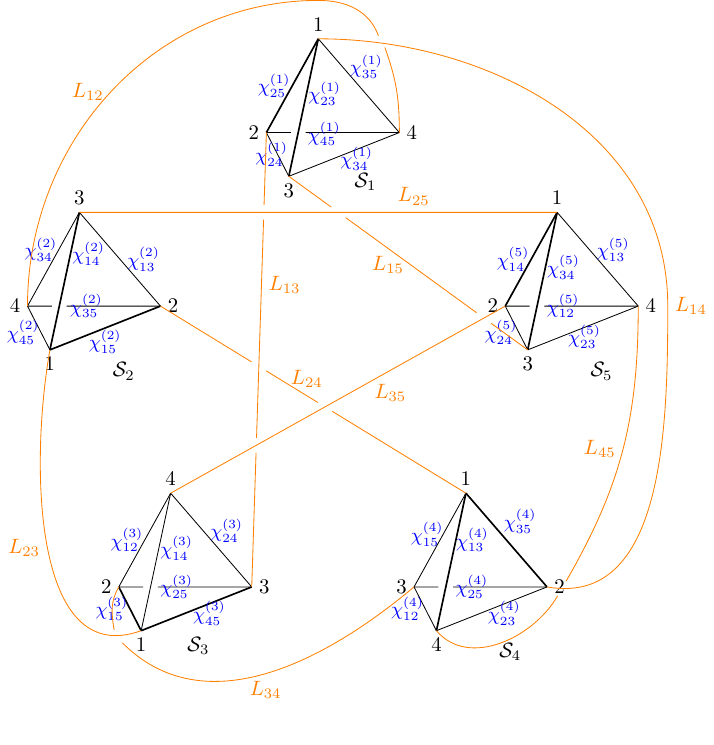}
\caption{Ideal triangulations $\{{\bf T}(\cS_a)\}$ of 4-holed spheres $\{\cS_a\}$ on $\partial(\SG)$. Each cusp boundary of ${\bf T}(\cS_a)$ is shrunk to a vertex, labeled by a number $i\in\{1,2,3,4\}$. The FG coordinates $\{\chi_{cd}^{(a)}\}$ of edges of ${\bf T}(\cS_a)$ are shown. Lines in orange represent the annuli connecting cusps of $\{{\bf T}(\cS_a)\}$ and the associated FN length on each annulus is specified. }
\label{fig:identify}
\end{figure}

Let the edge dressed with the FG coordinate $\chi_{cd}^{(a)}$ on $\cS_a$ be identified with the edge dressed with ${\chi'}^{(a)}_{ef}$ on $\cS'_a$. We choose the way of gluing so that the holes labelled by the same number ($1,\cdots, 4$) ({\it r.f.} fig.\ref{fig:identify}) are identified, which gives the constraints
\be
\chi^{(a)}_{cd}+{\chi'}^{(a)}_{cd}=2\pi i\,,\quad
\forall\, c,d\neq a\,.
\label{eq:glue_chi}
\ee

It is easy to verify that they are first-class constraints. These constraints can also be reformulated into ones written in terms of coordinates $\{\{L_{ac}\}_c,M_a,P_a\}$ and $\{\{L'_{ac}\}_c,M'_a,-P'_a\}$ after symplectic transformation by \eqref{eq:chi_to_LMP}. It gives four simple constraints, denoted as $\cC_1,\cdots,\cC_4$, on the FN lengths
\be
2L_{ac}+2L'_{ac}=0\,,\quad\forall\,c\neq a
\label{eq:glue_L}
\ee
pairwise. The pairing reflects the chosen gluing of holes from $\cS_a$ and $\cS'_a$. On the other hand, the remaining two constraints $\cC_5$ and $\cC_6$ are constraints on FG coordinates $\{M_a,P_a,M'_a,-P'_a\}$\footnote{In general, there are 12 ways to glue the holes from $\cS_a$ and $\cS'_a$ pairwise. Except for identifying the labels of holes, there is another gluing that also gives the same results of $\cC_5$ and $\cC_6$ as in \eqref{eq:cC_5-cC_6} but a different set of $\cC_{1,\cdots,4}$. For the remaining 10 ways of gluing, $\cC_5$ and $\cC_6$ are linear combinations of $\{L_{ac}\}_c$ and $\{M_a,P_a,M'_a,-P'_a\}$ with integer coefficents. }:
\be
M_a+M'_a=2\pi i\,,\quad 
P_a+P'_a=2\pi i\,.
\label{eq:cC_5-cC_6}
\ee 

These constraints are part of the position variables after a symplectic transformation on the coordinates on the union of two $\SG$'s. We define 
\be\begin{aligned}
\vec{\cQ}_{\rm ini}&=\{\vec{\cQ}_a,\vec{\cQ}'_a\}^\top\,,\quad
\text{with } \vec{\cQ}_a=\{2L_{ac},2L_{ad},2L_{ae},2L_{af},M_a\}^\top\,,
\vec{\cQ}'_a=\{2L'_{ac},2L'_{ad},2L'_{ae},2L'_{af},M'_a\}^\top\,,\\
\vec{\cP}_{\rm ini}&=\{\vec{\cP}_a,\vec{\cP}'_a\}^\top\,,\quad
\text{with } \vec{\cP}_a=\{T_{ac},T_{ad},T_{ae},T_{af},P_a\}^\top\,,
\vec{\cP}'_a=\{T'_{ac},T'_{ad},T'_{ae},T'_{af},P'_a\}^\top\,.
\end{aligned}
\label{eq:Q_P_fin}
\ee
They are transformed to 
\be
\vec{\cQ}_{\rm fin}=\{2L_{ac},2L_{ad},2L_{ae},2L_{af},\vec{\cC}\}^\top\,,\quad
\vec{\cP}_{\rm fin}=\{T_{ac}+T'_{ac},T_{ad}+T'_{ad},T_{ae}+T'_{ae},T_{af}+T'_{af},\vec{\Gamma}\}\,,
\label{eq:gauge_orbit_glue}
\ee
where $\vec{\Gamma}$ is the vector of gauge orbits which reads
\be
\vec{\Gamma}= \{T'_{ac},T'_{ad},T'_{ae},T'_{af},M'_a,-P'_a\}^\top\,.
\ee
The initial and final coordinates are related by a symplectic transformation that can be separated into a U-type, a partial S-type and an affine translation. Explicitly,
\be
\mat{c}{\vec{\cQ}_{\rm fin}\\\vec{\cP}_{\rm fin}}
={\bf M}_{aa}
\mat{c}{\vec{\cQ}_{\rm ini}\\\vec{\cP}_{\rm ini}} + i\pi \vec{t}_{aa}\,,\quad
{\bf M}_{aa}={\bf S}_{10}\cdot{\bf U}({\bf A}_{aa})\,,\quad \text{with }
{\bf A}_{aa}=\left(\begin{array}{cccccccccc}
 1 & 0 & 0 & 0 & 0 & 0 & 0 & 0 & 0 & 0 \\
 0 & 1 & 0 & 0 & 0 & 0 & 0 & 0 & 0 & 0 \\
 0 & 0 & 1 & 0 & 0 & 0 & 0 & 0 & 0 & 0 \\
 0 & 0 & 0 & 1 & 0 & 0 & 0 & 0 & 0 & 0 \\
 1 & 0 & 0 & 0 & 0 & 1 & 0 & 0 & 0 & 0 \\
 0 & 1 & 0 & 0 & 0 & 0 & 1 & 0 & 0 & 0 \\
 0 & 0 & 1 & 0 & 0 & 0 & 0 & 1 & 0 & 0 \\
 0 & 0 & 0 & 1 & 0 & 0 & 0 & 1 & 0 & 0 \\
 0 & 0 & 0 & 0 & 1 & 0 & 0 & 0 & 0 & 1 \\
 0 & 0 & 0 & 0 & 0 & 0 & 0 & 0 & 0 & 1 \\
\end{array}
\right)\,,
\ee
and only the 9-th and 10-th elements of $\vec{t}_{aa}$:
\be
\lb\vec{t}_{aa}\rb_{9}=\lb\vec{t}_{aa}\rb_{10}=-2\,,\quad
\lb\vec{t}_{aa}\rb_i=0,\,\forall\, i\neq 9,10\,.
\ee
When one considers the full set of coordinates on two $\SG$'s glued through 4-holed spheres, ${\bf M}_{aa}$ can be seen as a submatrix of a larger symplectic matrix.  

To impose the (quantum) constraints on the partition function, we first perform the Weil transformation on the product of two partition functions, each on one of the $\SG$'s, as we did in Section \ref{subsec:CS_partition}. 
That is, 
\be
\cZ^0_{\rm fin}\lb \vec{\mu}|\vec{m}\rb
=\f{1}{k}\sum_{n_{10}\in4(\Z/2N\Z)} \int\rd\nu_{10} 
e^{\f{2\pi i}{k}\left[-\lb\mu_{10}+iQ\rb\nu_{10}+m_{10}n_{10}\right]}
\cZ^k_{\rm ini}\lb\bA_{aa}^{-1}(\mu_{1,\cdots, 8},\mu_9+iQ,\nu_{10})^\top|\bA_{aa}^{-1}(m_{1,\cdots, 9},n_{10})^\top\rb\,, 
\label{eq:Z_fin}
\ee
where $\vec{\mu},\vec{m}$ are parameters of $\vec{\fQ}_{\rm fin}$ and $\cZ^k_{\rm ini}$ is the product of two $\cZ_{\SG}$'s defined as in \eqref{eq:SG_partition_final} when its discrete variables are in $4(\Z/2N\Z)$ while defined trivially otherwise, similar to the definition \eqref{eq:def_Z2} (we have omitted the variables irrelevant to the gluing process). $\bA^{-1}_{aa}$ is an integer matrix. Parametrize
\be
2L_{ac}=\f{2\pi i}{k}\lb -i b \mu_{ac}-m_{ac}\rb\,,\quad 
2L'_{ac}=\f{2\pi i}{k}\lb -i b \mu'_{ac}-m'_{ac}\rb\,,\quad\forall\, c\neq a\,,\quad
\cC_i = \f{2\pi i}{k}\lb -i b \mu_{\cC_i}-m_{\cC_i}\rb\,,\quad \forall\, i=1,\cdots,6\,.
\ee
Then $\mu_{\cC_{1,\cdots,4}}=\{\mu_{ac}-\mu'_{ac}\}_c,\, \mu_{\cC_5}=\mu_a+\mu'_a-iQ\equiv \mu_9,\,\mu_{\cC_6}=\nu_a+\nu'_a-iQ\equiv \mu_{10}$ and $m_{\cC_{1,\cdots,4}}=\{m_{ac}-m'_{ac}\}_c,\, m_{\cC_5}=m_a+m'_a,\,m_{\cC_6}=n_a+n'_a\equiv m_{10}$. 
The constraint $\cC_i=0$ is quantized to $\widehat{\cC}_i=\hbar\,,\forall\, i=1,\cdots,6$, imposing which on \eqref{eq:Z_fin} gives 
\be
\cZ^k_{\rm fin}\lb \vec{\mu}|\vec{m}\rb
=\f{1}{k}\sum_{m_a\in4(\Z/2N\Z)} \int\rd\mu_{a} 
e^{-\f{2\pi }{k}Q}
\cZ_{\SG}(\{\mu_{ac}\}_c,\mu_a|\{m_{ac}\}_c,m_a) \cZ_{\SG}(\{-\mu_{ac}\}_c,iQ-\mu_a|\{-m_{ac}\}_c,-m_a)\,.
\label{eq:Z_fin_constraint}
\ee
Note that $\cC_6=0$ implies $\beta_a+\beta'_a\equiv\im(\nu_a)+\im(\nu'_a)=Q$. Then \eqref{eq:Z_fin_constraint} can be rewritten as
\begin{multline}
\cZ^k_{\rm fin}\lb \vec{\mu}|\vec{m}\rb
= \f{1}{k}\sum_{m_a\in4(\Z/2N\Z)} \int_\R\rd\mu_{a} 
\lb e^{-\f{2\pi \beta_a}{k}}
\cZ_{\SG}(\{\mu_{ac}\}_c,\mu_a+i\alpha_a|\{m_{ac}\}_c,m_a)\rb\\
\lb e^{-\f{2\pi \beta'_a}{k}} \cZ_{\SG}(\{-\mu_{ac}\}_c,i(Q-\alpha_a)-\mu_a|\{-m_{ac}\}_c,-m_a)\rb\,,
\end{multline}
where the integrand is the product of two Schwarz functions on the integration contour \cite{Han:2023hbe}. This manifests that $\cZ^k_{\rm fin}$ is absolutely convergent. 

We also insert a pair of coherent states to $\cS_a$ and $\cS'_a$ respectively by expanding the delta distribution as
\be
\delta_{\mu_a^{\prime},-\mu_a} \delta_{e^{\frac{2 \pi i}{k}\left(m_a+m_a^{\prime}\right)}, 1}=\left(\frac{k}{4 \pi^2}\right)^2 \int_{\bC \times \bT^2} \mathrm{~d} \rho_a \Psi_{\rho_a}\left(\mu_a \mid m_a\right) \bar{\Psi}_{\rho_a}\left(-\mu_a^{\prime} \mid-m_a^{\prime}\right) \equiv \int_{\bC \times \bT^2} \mathrm{~d} \rho_a \Psi_{\rho_a}\left(\mu_a \mid m_a\right) \Psi_{\tilde{\rho}_a}\left(\mu_a^{\prime} \mid m_a^{\prime}\right)\,,
\ee
where $\tilde{\rho}_a=(-\bar{z}_a,-x_a,y_a)$ given $\rho_a=(z_a,x_a,y_a)$. We have used the property of the coherent state: $\Psi_{\tilde{\rho}_a}(\mu|m)=\bar{\Psi}_{\rho_a}(-\mu|-m)$. 
Coupling with coherent states to unglued 4-holed spheres and imposing the simplicity constraints on all the coherent states, one obtains the amplitude for the 4-complex, denoted as $\boldsymbol{\sigma}_{aa}$, composed by gluing two 4-simplices through a pair of tetrahedra, which can be separated into the integral of two vertex amplitudes:
\be
\cA_{\boldsymbol{\sigma}_{aa}}=
\frac{k}{\left(4 \pi^2\right)^2} \int_{\overline{\cM}_{\vec{j}^v_a}} \mathrm{~d} \hat{\rho}_a \lb e^{-\f{2\pi \beta_a}{k}}\cA_{v}(\iota)\rb \lb e^{-\f{2\pi \beta'_a}{k}}\cA_{v'}(\iota')\rb\,.
\label{eq:2-vertex_amplitude}
\ee
Here, $\iota$ and $\iota'$ encode the (correlated) data relevant to $\cS_a$ and $\cS'_a$, $\cA_{v}(\iota)$ and $\cA_{v'}(\iota')$ are constructed by coupling with constrainted coherent states $\Psi_{\hat{\rho}_a}\left(\mu_a \mid m_a\right) $ and $\Psi_{\hat{\tilde{\rho}}_a}\left(\mu_a^{\prime} \mid m_a^{\prime}\right)$ respectively and the integration $\overline{\cM}_{\vec{j}^v_a}$ is over a compact space of the curved tetrahedron shapes given triangle areas fixed by spins (see \cite{Han:2023hbe} for more details). It was shown in \cite{Han:2023hbe} (see Lemma III.1 therein) that $\left|\lb e^{-\f{2\pi \beta_a}{k}}\cA_{v}(\iota)\rb\right|$ and $\left|\lb e^{-\f{2\pi \beta'_a}{k}}\cA_{v'}(\iota')\rb\right|$ are both bounded from above on $\bC\times\bT^2$ for any given boundary data. Therefore, $\cA_{\boldsymbol{\sigma}_{aa}}$ is a finite amplitude. 

We can also rewrite \eqref{eq:2-vertex_amplitude} in a way that respects the {\it local amplitude ansatz} by introducing an edge amplitude.
We define the edge amplitude associated to the spinfoam edge $e$ resulting from gluing $\cS_a$ from spinfoam vertex $v$ and $\cS_a'$ from spinfoam vertex $v'$ to be a set of delta distributions on the relevant spins, coherent state labels and positive angles:
\be
\cA_e^{vv'}\lb\iota^{vv'}_{a}\rb 
:= \f{k}{(2\pi)^4}\delta_{\vec{j}_a^{v}-\vec{j}_a^{v'}}
\delta_{\hat{\rho}_a'-\hat{\tilde{\rho}}_a}\delta_{\alpha_a'+\alpha_a} e^{-\f{2\pi (\beta_a+\beta'_a)}{k}}\,,
\label{eq:edge_amplitude}
\ee
where $\iota^{vv'}_{a}=\lb\vec{j}_a^{v},\vec{j}_a^{v'}, \hat{\rho}_a,\hat{\rho}_a',\alpha_a,\beta_a,\alpha_a',\beta_a'\rb$. 
Then we can rewrite
\be
\cA_{\boldsymbol{\sigma}_{aa}}=\int_{\overline{\cM}_{\vec{j}^v_a}} \mathrm{~d} \hat{\rho}_a \int_{\overline{\cM}_{\vec{j}^{v'}_a}} \mathrm{~d} \hat{\rho}'_a\, \cA_{v}(\iota)\cA_e^{vv'}(\iota^{vv'}_{a})\cA_{v'}(\iota')\,.
\ee

The same gluing process can be repeated to generate the spinfoam amplitude for any 4-complex generated by the colored tensor model. 

\medskip

To form the full amplitude, one also needs to add a face amplitude for each internal spinfoam face, if it exists. A spinfoam face $f$ corresponds to a torus on the boundary of the graph complement, to which a pair of FN length (fixed by a spin $j_f$) and FN twist is associated. We propose a simpler face amplitude compared to \cite{Han:2023hbe,Han:2024reo} with an undetermined power $\fp\in\R$ \footnotemark{}:
\be
\cA_f(2j_f):=[2j_f+1]_\fq^\fp \,,\quad\fp\in \R\,,\quad
j_f=0,2,\cdots,4N-2\,,
\label{eq:face_amplitude}
\ee
where $[n]_\fq:=\f{\fq^n-{\fq}^{-n}}{\fq-\fq^{-1}}\equiv \sin\lb \f{2\pi n}{k} \rb/\sin\lb \f{2\pi}{k}\rb$ is a $\fq$-number with $\fq=e^{2\pi i/k}$ being a root-of-unity depending on the CS level $k$. 
\footnotetext{
Compared to \cite{Han:2023hbe,Han:2024reo}, no phase factor is added to the face amplitude because we have chosen a different definition of FN twists through a different symplectic matrix as in \eqref{eq:symp_transf}. The phase factor was introduced in the previous constructions to cancel the mismatch of FN twists and the deficit angle hinged by the internal triangle dual to the spinfoam face. In the current construction, however, each of these FN twists has a direct relation to the deficit angle hence no phase factor is needed. }

To summarize, the spinfoam amplitude for a {spinfoam 2-complex} consisting of $V$ spinfoam vertices, $E_{\In}$ internal spinfoam edges and $F_{\In}$ internal spinfoam faces takes the form 

\be
\cZ_{\vec{\hrho}_\partial}(\vec{\alpha}|\vec{j}_b)
=\sum_{{\rm even }\,j_f=0}^{4N-2}
\int_{\overline{\cM}_{\vec{j}^v_a}}\rd \hrho^{v\in e}_a\int_{\overline{\cM}_{\vec{j}^{v'}_a}}\rd \hrho^{v'\in e}_a 
\left[\prod_{f=1}^{F_{\In}}\mathcal{A}_f(2j_f)\right]
\left[\prod_{e=1}^{E_{\In}}
\cA_e(\hrho_a^{v\in e},\hrho_a^{v'\in e}|\{j_{ac}^{v\in e},j_{ac}^{v'\in e}\}_{c\neq a})\right]
\left[\prod_{v=1}^V\cA_{v}(\vec{\alpha}^v,\vec{j}^v,\vec{\hrho}^v)\right],
\label{eq:spinfoam_amplitude}
\ee
where $v\in e$ denotes that $v$ is at the (source or target) end of $e$, $\vec{\alpha}$ contains all the positive angles, $\vec{\hrho}_\partial$ contains all the coherent state labels on the boundary, the summations in $j_f$ are for all the internal spinfoam faces and the integrations over coherent state labels are for all the internal spinfoam edges.

\section{Critical points of the spinfoam amplitude}
\label{sec:critical_point}

In this section, we derive the critical points of the spinfoam amplitude to extract their geometric interpretation. Similar analysis using a different set of variables can be found in \cite{Han:2023hbe,Han:2024reo}.
It relies on looking into the large $k$-regime of the amplitude \eqref{eq:spinfoam_amplitude}, which allows us to extract the action in terms of the integration variables and find the equations of motion. 
We realize that some variables are decoupled into different vertex amplitudes. It will, therefore, simplify the analysis by first analyzing the critical points of the vertex amplitude \eqref{eq:vertex_amplitude}.

\subsection{Critical points of the vertex amplitude}
\label{subsec:critical_point_vertex}

Two key steps are to rewrite \eqref{eq:vertex_amplitude} in terms of the scaleless variable \wrt $k$ and to change the finite summations into integrations through Poisson resummation. In the former step, we convert components of $\{\mu_I,\nu_I,m_I,n_I\}_{I=1}^{15}$ into $\{\fQ_I,\tfQ_I,\fP_I,\tfP_I\}_{I=1}^{15}$ that do not scale with $k$ through
\begin{subequations}
\begin{align}
\mu=\f{kb\lb\fQ_I+\tfQ_I\rb}{2\pi(b^2+1)}\,,\quad &
m_I=\f{ik\lb\fQ_I-b^2\tfQ_I\rb}{2\pi(b^2+1)}\,,
\label{eq:m_mu-FN_FG}\\
\nu_I=\f{kb\lb\fP_I+\tfP_I\rb}{2\pi(b^2+1)}\,, \quad &
n_I=\f{ik\lb\fP_I-b^2\tfP_I\rb}{2\pi(b^2+1)}\,.
\label{eq:n_nu-FN_FG}
\end{align}
\label{eq:mn_munu-FN_FG}
\end{subequations}
There are two kinds of Poisson resummations  for any function 
\begin{subequations}
\begin{align}
\sum_{n \in 4(\mathbb{Z} / 2N \mathbb{Z})} f(n)&\equiv \sum_{r \in \mathbb{Z} / 2N \mathbb{Z}} f(4r)
=\frac{N}{ \pi} \sum_{p \in \mathbb{Z}} \int_{-\delta / 2N}^{2 \pi-\delta / 2N} \mathrm{~d} \mathcal{J} f\left(\frac{k}{2 \pi} \mathcal{J}\right) e^{i 2N p \mathcal{J}}\,,\quad\\
\sum_{m_0 \in 2(\mathbb{Z} /4 N \mathbb{Z})} f(m_0)&=\frac{k}{2 \pi} \sum_{q \in \mathbb{Z}} \int_{-\delta / k}^{2 \pi-\delta / k} \mathrm{~d} \mathcal{K} f\left(\frac{k}{2 \pi} \mathcal{K}\right) e^{i k q \mathcal{K}}\,,
\end{align}
\end{subequations}
where $\cJ=2\pi n/k$ and $\cK=2\pi m_0/k$. 
The vertex amplitude \eqref{eq:vertex_amplitude} can be written as

	\be
\cA_v(\vec{\alpha},\vec{j},\vec{\hrho})\xrightarrow{k\rightarrow\infty} \cN 
\sum_{\vec{p}\in\Z^{15}}\sum_{\vec{q}\in\Z^{15}}\sum_{\vec{u}\in\Z^5}
\int_{\cC^{\times70}_{\cM}}\rd \cM\,
\exp\left[ k
S^{v}_{\vec{p},\vec{q},\vec{u},\vec{\hrho}}
(\vec{\fP},\vec{\tfP},\vec{\fQ}^v,\vec{\tfQ},\vec{\Phi},\vec{\widetilde{\Phi}})
\right]\left[1+O(1/k)\right]\,,
\label{eq:vertex_large_k}
\ee

where $\cN=\frac{i}{2^{42}}\left(\frac{N}{\pi^{2}Q}\right)^{30}\lb\f{2}{k}\rb^{\f52}=\frac{iN^{55/2}}{2^{47}\pi^{60}Q^{30}}$ and the measure reads
\be
\int\limits_{\cC^{\times70}_{\cM}}\rd \cM:=
\int\limits_{\cC^{\times 30}_{\fP\times\tfP}}
\bigwedge\limits_{I=1}^{15}\left(-i\, \mathrm{d} \mathfrak{P}_{I} \wedge \mathrm{d} \widetilde{\mathfrak{P}}_{I}\right)
\int\limits_{\cC^{\times 30}_{\Phi\times\tilde{\Phi}}}
\bigwedge\limits_{I=1}^{15}\left(-i\, \mathrm{d} \Phi_{I} \wedge \mathrm{d} \widetilde{\Phi}_{I}\right)
\int\limits_{\cC^{\times10}_{M_{a}\times\widetilde{M}_{a}}}
\bigwedge\limits_{a=1}^{5}\lb-i\, \mathrm{d} {M}_{a} \wedge \mathrm{d} \widetilde{M}_{a}\rb\,.
\ee
The action in \eqref{eq:vertex_amplitude} can be separated into several parts:

\begin{multline}
S^{v}_{\vec{p},\vec{q},\vec{u},\vec{\hrho}}
=S_{0}\left(\mathfrak{\vec{Q}},\vec{\widetilde{\mathfrak{Q}}}\right)
+S_{1}\left(\mathfrak{\vec{Q}},\vec{\widetilde{\mathfrak{Q}}},\mathfrak{\vec{P}},\vec{\widetilde{\mathfrak{P}}},\vec{\Phi},\vec{\widetilde{\Phi}}\right)
+S_{2}\left(\vec{\Phi}\right)
+\widetilde{S}_{2}\left(\vec{\widetilde{\Phi}}\right)
+S_{3}\left(\mathfrak{\vec{P}},\vec{\widetilde{\mathfrak{P}}}\right)\\
+\sum\limits_{a=1}^5
\left[S_{\zh_{a}}(M_{a},\widetilde{M}_{a})
+S_{(\xh_{a},\yh_{a})}(M_{a},\widetilde{M}_{a})-\frac{\left(M_a-b^{2}\widetilde{M}_a\right)}{\left(b^{2}+1\right)}u_a\right]
-\frac{\left(\mathfrak{\vec{P}}-b^{2}\vec{\widetilde{\mathfrak{P}}}\right)}{4\left(b^{2}+1\right)}\cdot\vec{p}-\frac{\left(\vec{\Phi}-b^{2}\vec{\widetilde{\Phi}}\right)}{\left(b^{2}+1\right)}\cdot\vec{q}.
\label{eq:effective_action_all}
\end{multline}

The vectors $\vec{p},\vec{q}\in\Z^{15}$ and $\vec{u}\in\Z^5$ come from the Poisson resummations of $\vec{n}=\f{ik}{2\pi(b^2+1)}\lb\vec{\fP}-b^2\vec{\tfP}\rb$, $\vec{m}_0=\f{ik}{2\pi(b^2+1)}\lb\vec{\Phi}-b^2\vec{\widetilde{\Phi}}\rb$ ({\it r.f.} \eqref{eq:SG_partition_final}) and $m_a=\f{ik}{2\pi(b^2+1)}\lb M_a-b^2\widetilde{M}_a \rb, a=1,\cdots,5$ respectively.  
Neglecting the subleading contributions at large $k$, the first five terms on the {\it r.h.s.} of \eqref{eq:effective_action_all} are explicitly 

\begin{subequations}
\begin{align}
S_{0}\left(\mathfrak{\vec{Q}},\vec{\widetilde{\mathfrak{Q}}}\right)=&
\frac{i}{4\pi\left(b^{2}+1\right)^{2}}\left[-
\left(\vec{\mathfrak{Q}}-b^{2}\vec{\widetilde{\mathfrak{Q}}}\right)\cdot{\bf CA^{-1}}\cdot\left(\vec{\mathfrak{Q}}-b^{2}\vec{\widetilde{\mathfrak{Q}}}\right)-b^{2}\left(\vec{\mathfrak{Q}}+\vec{\widetilde{\mathfrak{Q}}}\right)\cdot{\bf CA^{-1}}\cdot\left(\vec{\mathfrak{Q}}+\vec{\widetilde{\mathfrak{Q}}}\right)\right] \\
\label{eq:S0}\\
S_{1}\left(\mathfrak{\vec{Q}},\vec{\widetilde{\mathfrak{Q}}},\mathfrak{\vec{P}},\vec{\widetilde{\mathfrak{P}}},\vec{\Phi},\vec{\widetilde{\Phi}}\right)=&
\frac{i}{2\pi\left(b^{2}+1\right)^{2}}\left[b^{2}
\left({\bf A}\left(\vec{\Phi}+\vec{\widetilde{\Phi}}\right)-\left(\vec{\mathfrak{Q}}+\vec{\widetilde{\mathfrak{Q}}}\right)\right)\cdot\left(\vec{\mathfrak{P}}+\vec{\widetilde{\mathfrak{P}}}\right)\right.\nn\\
&+\left.\left({\bf A}\cdot\left(\vec{\Phi}-b^{2}\vec{\widetilde{\Phi}}\right)-\left(\vec{\mathfrak{Q}}-b^{2}\vec{\widetilde{\mathfrak{Q}}}\right)
\right)\cdot\left(\vec{\mathfrak{P}}-b^{2}\vec{\widetilde{\mathfrak{P}}}\right)\right]
\label{eq:S1}\\
S_{2}\left(\vec{\Phi}\right)=&-\frac{i}{2\pi\left(b^{2}+1\right)}\sum_{i=1}^{5}\left[{\rm Li}_{2}\left(e^{-X_{i}}\right)+\mathrm{Li}_{2}\left(e^{-Y_{i}}\right)+\mathrm{Li}_{2}\left(e^{-Z_{i}}\right)+\mathrm{Li}_{2}\left(e^{-W_{i}}\right)\right]\,,
\label{eq:S2}\\
\widetilde{S}_{2}\left(\vec{\widetilde{\Phi}}\right)=&-\frac{i}{2\pi\left(b^{-2}+1\right)}\sum_{i=1}\left[{\rm Li}_{2}\left(e^{-\tilde{X}_{a}}\right)+{\rm Li}_{2}\left(e^{-\tilde{Y}_{i}}\right)+{\rm Li}_{2}\left(e^{-\widetilde{Z}_{i}}\right)+{\rm Li}_{2}\left(e^{-\widetilde{W}_{i}}\right)\right]\,,
\label{eq:S2t}\\
S_{3}\left(\mathfrak{\vec{P}},\vec{\widetilde{\mathfrak{P}}}\right)=&
\frac{i}{4\pi\left(b^{2}+1\right)^{2}}\left[\left(\mathfrak{\vec{P}}-b^{2}\vec{\widetilde{\mathfrak{P}}}\right)\cdot{\bf BA^{\top}}\cdot\left(\mathfrak{\vec{P}}-b^{2}\vec{\widetilde{\mathfrak{P}}}\right)+b^{2}\left(\mathfrak{\vec{P}}+\vec{\widetilde{\mathfrak{P}}}\right)\cdot{\bf BA^{\top}}\cdot\left(\mathfrak{\vec{P}}+\vec{\widetilde{\mathfrak{P}}}\right)\right]\,,
\end{align}
\label{eq:S0-S1t}
\end{subequations}
where $\vec{\Phi}=\left(X_{i},Y_{i},Z_{i}\right)_{i=1}^{5}$ and $\vec{\widetilde{\Phi}}=\left(\tilde{X}_{i},\tilde{Y}_{i},\tilde{Z}_{i}\right)_{i=1}^{5}$
 with subscript $i$ denoting the octahedron $\Oct(i)$ on the $\SG$. Similarly for the tilde sectors. 
 The second line of \eqref{eq:S0} are subleading terms hence can be neglected in the stationary point analysis. 
 $\Li_2$ appearing in \eqref{eq:S2} and \eqref{eq:S2t} is the dilogarithm function defined as $\Li_2(z):=-\int_0^z\f{\ln(1-u)}{u}\rd u$ for $z\in\bC$.
The first two actions in the square bracket of \eqref{eq:effective_action_all} correspond to the coherent states \eqref{eq:coherent_state_1} and \eqref{eq:coherent_state_2} respectively and, neglecting subleading contributions, they read

\begin{subequations}
\begin{align}
S_{\zh_{a}}(M_{a},\widetilde{M}_{a})&=
-\f{ b}{2\pi(b^2+1)}\lb M_{a}+\widetilde{M}_a\rb
\left[\f{b\lb M_{a}+\widetilde{M}_{a}\rb}{2(b^2+1)}-\sqrt{2}\hat{\bar{z}}_{a}\right]-\f{1}{2\pi}\re(\zh_{a})^2\,,
\label{eq:coherent_QQt_1}\\
S_{(\xh_{a},\yh_{a})}(M_{a},\widetilde{M}_{a})
&=- \f{i\xh_{a}\yh_{a}}{4\pi}-\f{1}{4\pi}\left[\f{i\lb M_{a}-b^2\widetilde{M}_{a}\rb}{ b^2+1}-\xh_{a}\right]^2-\f{1}{2\pi}\f{\lb M_{a}-b^2\widetilde{M}_{a}\rb  \yh_{a}}{b^2+1} 
\,.
\label{eq:coherent_QQt_2}
\end{align}
\label{eq:coherent_QQt}
\end{subequations}

We analyze the equations of motion of different parts of the action \wrt the integration variables $\{\mathfrak{P}_{I},\mathfrak{\tilde{P}}_{I},\Phi_{I},\tilde{\Phi}_{I}\}$:
\begin{subequations}
\begin{align}
-2\pi i\left(b^{2}+1\right)\frac{\partial S_{1}}{\partial\mathfrak{P}_{I}}
&=\left({\bf A}_{IJ}\Phi_{J}-\mathfrak{Q}_{I}
\right)\,,\quad
-2\pi i\left(b^{-2}+1\right)\frac{\partial S_{1}}{\partial\mathfrak{\tilde{P}}_{I}}
=2\left({\bf A}_{IJ}\tilde{\Phi}_{J}-\widetilde{\mathfrak{Q}}_{I}
\right)\,,\\
-2\pi i\left(b^{2}+1\right)\frac{\partial S_{1}}{\partial\Phi_{I}}&=
{\bf A}_{IJ}^{\top}\mathfrak{P}_{J}\,,\quad
-2\pi i\left(b^{-2}+1\right)\frac{\partial S_{1}}{\partial\tilde{\Phi}_{I}}=
{\bf A}_{IJ}^{\top}\widetilde{\mathfrak{P}}_{J}\,,\\
-2\pi i\left(b^{2}+1\right)\frac{\partial S_{2}}{\partial \Phi_{I}}
&=-P_{I}\,,\quad
-2\pi i\left(b^{-2}+1\right)\frac{\partial \widetilde{S}_{2}}{\partial\tilde{\Phi}_I}=-\tilde{P}_{I}\,,\\
-2\pi i\left(b^{2}+1\right)\frac{\partial S_{3}}{\partial\mathfrak{P}_{I}}&=
{\bf \left(BA^{\top}\right)}_{IJ}\mathfrak{P}_{J}\,,\quad
-2\pi i\left(b^{-2}+1\right)\frac{\partial S_{3}}{\partial\mathfrak{\tilde{P}}_{I}}=
{\bf \left(BA^{\top}\right)}_{IJ}\mathfrak{\tilde{P}}_{J}\,,
\end{align}
\end{subequations}
where
\be\begin{aligned}
\vec{P}&=\left({\rm ln}\left(\frac{1-e^{-X_{i}}}{1-e^{-W_{i}}}\right),{\rm ln}\left(\frac{1-e^{-Y_{i}}}{1-e^{-W_{i}}}\right),{\rm ln}\left(\frac{1-e^{-Z_{i}}}{1-e^{-W_{i}}}\right)\right)_{i=1}^{5},\qquad\\
\vec{\tilde{P}}&=\left({\rm ln}\left(\frac{1-e^{-\tilde{X}_{i}}}{1-e^{-\tilde{W}_{i}}}\right),{\rm ln}\left(\frac{1-e^{-\tilde{Y}_{i}}}{1-e^{-\tilde{W}_{i}}}\right),{\rm ln}\left(\frac{1-e^{-\tilde{Z}_{i}}}{1-e^{-\tilde{W}_{i}}}\right)\right)_{i=1}^{5}\,.
\end{aligned}
\ee
Combinatorially, part of the critical equations of $S_{\rm tot}\equiv S^{v}_{\vec{p},\vec{q},\vec{u},\vec{\hrho}}$ gives
\begin{subequations}
\begin{align}
	-2\pi i\left(b^{2}+1\right)\frac{\partial S_{{\rm tot}}}{\partial\mathfrak{P}_{I}}	
	&={\bf A}_{IJ}\Phi_{J}-\mathfrak{Q}_{I}+{\bf \left(BA^{\top}\right)}_{IJ}\mathfrak{P}_{J}
	+\pi ip_{I}/2
	=0\,,
\label{eq:eom_P}\\
-2\pi i\left(b^{-2}+1\right)\frac{\partial S_{{\rm tot}}}{\partial\mathfrak{\tilde{P}}_{I}}	
&={\bf A}_{IJ}\tilde{\Phi}_{J}-\widetilde{\mathfrak{Q}}_{I}+{\bf \left(BA^{\top}\right)}_{IJ}\mathfrak{\tilde{P}}_{J}
-\pi ip_{I}/2
=0\,,
\label{eq:eom_Pt}\\
-2\pi i\left(b^{2}+1\right)\frac{\partial S_{{\rm tot}}}{\partial\Phi_{I}}	
&={\bf A}_{IJ}^{\top}\mathfrak{P}_{J}-P_{I}+2\pi iq_{I}=0\,,
\label{eq:eom_Phi}\\
-2\pi i\left(b^{-2}+1\right)\frac{\partial S_{{\rm tot}}}{\partial\tilde{\Phi}_{I}}	
&={\bf A}_{IJ}^{\top}\widetilde{\mathfrak{P}}_{J}-\tilde{P}_{I}-2\pi iq_{I}=0\,.
\label{eq:eom_Phit}
\end{align}
\end{subequations}
Notice that $\vec{\fP}$ and $\vec{\fQ}$ are obtained after three Weil transformations according to the symplectic transformation
\be
\mat{c}{\vec{\fQ}\\\vec{\fP}}=\mat{cc}{0 & -\id \\ \id & 0}
\mat{cc}{\id & 0\\-\bB\bA^\top & \id}
\mat{cc}{(\bA^{-1})^\top & 0 \\ 0 & \bA}
\mat{cc}{0 & \id \\ -\id & 0}\mat{c}{\vec{\Phi}\\\vec{\Pi}}
=
\mat{cc}{\bA  & \bB\\ 0 &(\bA^{-1})^\top}\mat{c}{\vec{\Phi}\\\vec{\Pi}}\,,
\label{eq:PQ_from_PhiPi}
\ee
which gives
\be
\vec{\Pi}=\bA^\top \cdot \vec{\fP}\,,\quad 
\label{eq:Pi_to_P}
\ee
Similarly for the tilde sectors.
Then \eqref{eq:eom_Phi} and \eqref{eq:eom_Phit} leads to $P_I-\Pi_I=\tilde{P}_I-\tilde{\Pi}_I=0$ and $q_I=0$, which is nothing but the reformulation of the algebraic curve equations (\ie a series of $z^{-1}+z''-1=0$'s) defining the moduli space of flat connection on $\SG$. 
Another equality from \eqref{eq:PQ_from_PhiPi} is
\be
\vec{\fQ}=\bA\cdot\vec{\Phi}+\bB\cdot\vec{\Pi}\,.
\label{eq:QtoPhiPi}
\ee
Plugging \eqref{eq:Pi_to_P} and \eqref{eq:QtoPhiPi} to \eqref{eq:eom_P} leads to 
\be
-\fQ_I+\fQ_I+\bB_{IJ}\Pi_J - \bB_{IJ} \Pi +i\pi
p_I/2=0\,,
\ee
which fixes $p_I=0$. 

We finally analyze the equations of motion of $S_{\rm tot}$ \wrt the FG coordinates $\{M_a, \widetilde{M}_a\}_{a=1}^5$, which are composed by parts as follows. 
Notice that the derivatives of $S_0$ and $S_1$ are the same as those \wrt $\{M'_a\equiv M_a-i\pi t_{\alpha,a+10},\widetilde{M}'_a\equiv \widetilde{M}'_a+i\pi t_{\alpha,a+10}\}_{a=1}^5$. Therefore, 
\begin{subequations}
\begin{align}
-2\pi i\left(b^{2}+1\right)\frac{\partial S_{0}}{\partial M_a}&=
-\left({\bf CA^{-1}}\right)_{a+10,J}\mathfrak{Q}_{J}\,,\quad
-2\pi i\left(b^{2}+1\right)\frac{\partial S_{1}}{\partial M_a}=-\mathfrak{P}_{a+10}\,,
\label{eq:eom_M_1}\\
-2\pi i\left(b^{-2}+1\right)\frac{\partial S_{0}}{\partial\widetilde{M}_{a}}&=
-\left({\bf CA^{-1}}\right)_{a+10,J}\mathfrak{\tilde{Q}}_{J}
\,,\quad
-2\pi i\left(b^{-2}+1\right)\frac{\partial S_{1}}{\partial\widetilde{M}_{a}}=-\widetilde{\mathfrak{P}}_{a+10}\,.
\label{eq:eom_M_2}\\
-2\pi i\left(b^{2}+1\right)\frac{\partial S_{z_{a}}}{\partial M_{a}}&=ib\left(\frac{b\left(M_{a}+\widetilde{M}'_{a}\right)}{\left(b^{2}+1\right)}-\sqrt{2}\hat{\bar{z}}_{a}\right)\,,\\
-2\pi i\left(b^{-2}+1\right)\frac{\partial S_{z_{a}}}{\partial\widetilde{M}_{a}}&=-ib^{-1}\left(\frac{b^{-1}\left(M_{a}+\widetilde{M}_{a}\right)}{\left(b^{-2}+1\right)}-\sqrt{2}\hat{\bar{z}}_{a}\right)\,,\\
-2\pi i\left(b^{2}+1\right)\frac{\partial S_{(x_{a},y_{a})}}{\partial M_{a}}&=-i\left(\frac{M_{a}-b^{2}\widetilde{M}_{a}}{\left(b^{2}+1\right)}+i\left(\hat{x}_{a}+i\hat{y}_{a}\right)\right)\,,\\
-2\pi i\left(b^{-2}+1\right)\frac{\partial S_{(x_{a},y_{a})}}{\partial\widetilde{M}_{a}}&=ib^{-2}\left(\frac{M_{a}-b^{2}\widetilde{M}_{a}}{\left(b^{-2}+1\right)}+i\left(\hat{x}_{a}+i\hat{y}_{a}\right)\right)
\end{align}	
\label{eq:eom_M}
\end{subequations}
Note that the final coordinates before the affine translations, denoted as $\lb\vec{\fQ}',\vec{\fP}' \rb$, after all the symplectic transformations are related to $\lb\vec{\fQ},\vec{\fP}\rb$ by
\be
\mat{c}{\vec{\fQ}'\equiv\vec{\cQ}-i\pi\vec{t}_\alpha\\\vec{\fP}'\equiv\vec{\cP}-i\pi\vec{t}_\beta }=\mat{cc}{\id & 0 \\ {\bf CA^{-1}} & \id}\mat{c}{\vec{\fQ}\\\vec{\fP}}\,.
\ee
Therefore, the FG coordinate $P_a$ conjugate to $M_a$ are
\be
P_a= \sum_{J}\lb{\bf CA^{-1}}\rb_{a+10,J} \fQ_J +\fP_{a+10}\,,\quad \forall\, a=1,\cdots,5\,.
\ee
The same holds for the tilde sector. 
Using this fact and combinig the equations in \eqref{eq:eom_M} as well as the derivative of the term $-\frac{\left(M_a-b^{2}\widetilde{M}_a\right)}{b^{2}+1}u_a$ in \eqref{eq:effective_action_all}, we get the equations of motion
\begin{subequations}
\begin{align}
-2\pi i\left(b^{2}+1\right)\frac{\partial S_{\rm tot}}{\partial M_{a}}
&=2\pi iu_a
-P_{a}+\frac{2\pi ib}{k}\left(\mu_{a}-\frac{k}{\sqrt{2}\pi}\hat{\bar{z}}_{a}\right)-\frac{2\pi}{k}\left(m_{a}-\frac{k}{2\pi}\left(\hat{x}_{a}+i\hat{y}_{a}\right)\right)=0\,,
\label{eq:eom_total_M_1}\\
-2\pi i\left(b^{-2}+1\right)\frac{\partial S_{\rm tot}}{\partial\widetilde{M}_{a}}
&=-2\pi iu_a
-\widetilde{P}_{a}+\frac{2\pi ib^{-1}}{k}\left(\mu_{a}-\frac{k}{\sqrt{2}\pi}\hat{\bar{z}}_{a}\right)+\frac{2\pi}{k}\left(m_{a}-\frac{k}{2\pi}\left(\hat{x}_{a}+i\hat{y}_{a}\right)\right)=0\,,
\label{eq:eom_total_M_2}
\end{align}
\label{eq:eom_total_M}
\end{subequations}
the solution to which matches the peak \eqref{eq:eom_position} of the coherent states $\Psi_{\hrho}(\mu_a|m_a)$:
\be
u_a=0\,,\quad
\mu_{a}=\frac{k}{\sqrt{2}\pi}{\rm Re}(z_{a})\,,\quad
\nu_{a}=-\frac{k}{\sqrt{2}\pi}{\rm Im}(z_{a})\,,\quad
m_{a}=\frac{k}{2\pi}\hat{x}_{a}\,,\quad
n_{a}=-\frac{k}{2\pi}\hat{y}_{a}\,.
\ee

\subsection{Critical points with respect to variables across 4-simplices}
\label{subsec:critical_point_other}

After a full analysis of the critical points for a single vertex amplitude. We now turn to analyze the critical points of the spinfoam amplitude \eqref{eq:spinfoam_amplitude} of a general spacetime triangulation that corresponds to a colorable spinfoam graph. 
Since the integration variables $\{\fP_I,\tfP_I,\Phi_I,\tilde{\Phi}_I\}_{I=1}^{15}$ as well as the FG coordinates $\{M_a,\widetilde{M}_a\}$ on the boundaries are decoupled in different vertex amplitudes, the revelant critical points above are still valid for the full amplitude. What remains to be derived are the critical points from the variation \wrt the coherent state labels appearing in the edge amplitudes $\cA_e$'s as well as the FN length $2L_f$'s appearing in the face amplitudes. 

We first look at the critical points from the variation \wrt to the coherent state labels shared by two 4-simplies. Consider the following part of the full amplitude: 
\be
\frac{k}{(2\pi)^4}\int_{\overline{\cM}_{\vec{j}_a}}\rd \hrho_a \,e^{-\frac{2\pi (\beta^L_a+\beta^R_a)}{k}}
 \cA_{v^L} (\vec{\alpha}^{v^L},\{j_{ac}\}_c,\{j^L_{bd}\}_{b,d\neq a},\hrho_a,\{\hrho^L_b\}_{b\neq a}) 
 \cA_{v^R} (\vec{\alpha}^{v^R},\{j_{ac}\}_c,\{j^R_{bd}\}_{b,d\neq a},\hat{\tilde{\rho}}_a,\{\hrho^R_b\}_{b\neq a})\,.
\ee
It corresponding to two glued 4-simplices through one shared tetrahedron, hence it includes two vertex amplitudes with shared data $\{j_{ac}\}_c$ and $\hrho_a$. The critical point can be found in the following action ({\it r.f. }\eqref{eq:coherent_QQt})
\be
S_{\zh_{a}}(M^L_{a},\widetilde{M}^L_{a}) + S_{-\hat{\bar{z}}_{a}}(M^R_{a},\widetilde{M}^R_{a})+
S_{(\xh_{a},\yh^v_{a})}(M^L_{a},\widetilde{M}^L_{a})+S_{(-\xh_{a},\yh^v_{a})}(M^R_{a},\widetilde{M}^R_{a})\,,
\ee
where $(M^L_{a},\widetilde{M}^L_{a})$ are the FG coordinates in $\cA_{v^L}$ while $(M^R_{a},\widetilde{M}^R_{a})$ in $\cA_{v^R}$. It is then easy to find that the critical points require these two pairs of FG coordinates to match up to a sign, \ie
\be
\mu^L_a=-\mu^R_a\,,\quad m^L_a=-m^R_a\,.
\ee

Lastly, we derive the critical points from the variation \wrt $2L_f$ associated to the triangle shared by, say, $|f|$ 4-simplices. The relevant part of the full amplitude is composed of a face amplitude and $|f|$ vertex amplitudes depending on $j_f$ (note that the first-class simplicity constraints restrict $\tilde{L}_f=-L_f$ hence only one integration variable is left):
\be
\sum_{j_f}[2j_f+1]_\fq^\fp \prod_{i=1}^{|f|} \cA_{v_i}(j_f,\cdots)=\sum_{u_f\in\Z}\int_{-\delta/k}^{2\pi-\delta/k}\rd (i2L_f)\,[2j_f+1]_\fq^\fp e^{-2ku_fL_f} \cA_{v_i}(j_f,\cdots)\,,
\ee 
where the ellipsis denotes the rest of the variables in the vertex amplitudes. The relevant action is as follows. ({\it r.f. }\eqref{eq:S0} for $S_0$ and \eqref{eq:S1} for $S_1$)
\be
S_{j_f}=\sum_{i=1}^{|f|} \left[S_0^{v_i}(2L_f,\cdots)+ S_1^{v_i}(2L_f,\cdots) \right]-2u_fL_f\,.
\ee
Similar to \eqref{eq:eom_M_1}, we get, for all $f=1,\cdots,10$,
\be\begin{split}
-2\pi i\left(b^{2}+1\right)\frac{\partial S_{j_f}}{\partial (2L_f)}
=&\sum_{i=1}^{|f|}\left[-\left({\bf CA^{-1}}\right)_{fJ}\mathfrak{Q}^{v_i}_{J}-\mathfrak{P}^{v_i}_{f}
+b^2\left({\bf CA^{-1}}\right)_{fJ}\widetilde{\mathfrak{Q}}^{v_i}_{J}
+b^2\widetilde{\mathfrak{P}}^{v_i}_{f}\right]+2\pi i\left(b^{2}+1\right)u_f\\
=&-\sum_{i=1}^{|f|}\left[\left({\bf CA^{-1}}\right)_{fJ}\lb \mathfrak{Q}^{v_i}_{J}-b^2\widetilde{\mathfrak{Q}}^{v_i}_{J}\rb +\lb \mathfrak{P}^{v_i}_{f}-b^2\widetilde{\mathfrak{P}}^{v_i}_{f}\rb\right]
  +2\pi i\left(b^{2}+1\right)u_f\\
\equiv& -\sum_{i=1}^{|f|} \lb T_f^{v_i}-b^2\widetilde{T}_f^{v_i}\rb +2\pi i \lb1+b^2\rb u_f =0\,.
\end{split}
\label{eq:eom_Lf}
\ee
As will be specified in the coming sections, the sum of the FN twists in the above equation is related to the {\it dressed} deficit angle $\varepsilon_f^{(s)}$ hinged by the internal triangle $f$, defined as the signed sum of dihedral angle hinged by $f$ in all adjacent 4-simplices $v$'s, \ie 
\be
\varepsilon_f^{(s)}:=\sum_{v \in f} s_v \Theta_f^v\,,\quad s_v=\operatorname{sgn}\left(V_4^v\right)\,.
\ee
The explicit relation of the summed FN twists and the dressed deficit angle is proven in \cite{Haggard:2014xoa} (see also Appendix B of \cite{Han:2024reo} for a summary):
\be
\cT_f:=\sum_{i=1}^{|f|} T_f^{v_i}=-\nu\varepsilon_f^{(s)}+2\pi i N_f\,,\quad
\widetilde{\cT}_f:=\sum_{i=1}^{|f|} \widetilde{T}_f^{v_i} = \nu\varepsilon_f^{(s)}-2\pi i N_f \,,\quad
N_f\in\Z\,,
\ee
where $N_f$ specifies the lift of the $\tau_f:=e^{\cT_f}$ to $\cT_f$. 
Then the solution to \eqref{eq:eom_Lf} gives a vanishing dressed deficit angle in the bulk as the critical geometry
\be
\varepsilon_f^{(s)} =0\,,\quad
u_f=N_f\,.
\ee
When $s_v={\rm cont}.$ for all 4-simplices sharing $f$, the deficit angle is zero. Since this is valid for all $f$ in the complex, the result implies that the critical point corresponds to the global dS$_4$ ($\nu>0$) or AdS$_4$ ($\nu>0$) geometry. 

\medskip

Let us finally summarize the critical points of the spinfoam amplitude \eqref{eq:spinfoam_amplitude}. Consider a 4-manifold $M_4$ whose spinfoam graph is colorable and whose triangulation gives 4-complex ${\bf T}(M_4)$ as a gluing of $V$ 4-simplices by identifying $E_{\rm in}$ pairs of tetrahedra. Denote the triangulation of its boundary $M_3=\partial M_4$ as ${\bf T}(M_3)$, and the dual graph of ${\bf T}(M_3)$ as $\Gamma$. The critical points of the amplitude are in a local coordinate patch\footnote{The symplectic manifold $\cM_\Flat(M_3\backslash \Gamma,\SL(2,\bC))$ is in general non-trivial and may contain singularities. To describe the full manifold, one needs to consider multiple coordinate patches to cover the whole manifold. In our construction, we only consider one coordinate patch. } of the {\it subset} of the moduli space of flat connection $\cM_\Flat(M_3\backslash \Gamma,\SL(2,\bC))$, denoted as $\cB_{\Gamma}$, which is formed by $V$ coordinate patches $\cB_{v}$'s each from one $\SG$ for a spinfoam vertex $v$, followed by imposing gluing constraints. They are related by
\be
\cB_{\Gamma} \equiv 
\lb\times_{v=1}^V \cB_{v}\rb/\left\{\vec{\cG}_e\right\}_{e=1}^{E_{\rm in}}\,,
\ee
where each vector $\vec{\cG}_e$ is a collection of 6 gluing constraints as in \eqref{eq:glue_L} and \eqref{eq:cC_5-cC_6} for gluing a pair of 4-holed spheres on the boundary of two $\SG$'s. We also denote real space of $\cB_{\Gamma}$, \ie the coordinate patch of $\cM_\Flat(M_3\backslash \Gamma,\SU(2)$, as $\cB_\Gamma^r$.

The subset is specified by the simplicity constraints. Firstly, the first-class simplicity constraints restrict the FN length part of the coordinates of $\cM_\Flat(\SG,\SL(2,\bC))$ to be within $\cM_\Flat(\SG,\SU(2))$. To specify the remaining restriction, we first realize that $\cM_\Flat(\SG,\SL(2,\bC))$ is a Lagrange submanifold of the symplectic space $\cM_\Flat(\partial(\SG),\SL(2,\bC))$, whose coordinate patch is denoted as $\cD_v$. 
Each admissible value of $L_{ab}=-4\pi i j_{ab}/k$ can be viewed as a gauge fixing of this symplectic space. 
One can obtain a symplectic subspace of it by the symplectic quotient of all the 10 gauges, which is a direct product of 5 phase spaces for 4-holed spheres. Denote the coordinate patch of $\cM_\Flat(\cS_a,\SL(2,\bC))$ as $\cD_e$. Then the symplectic quotient is,
\be
\cD_v//\{L_{ab}=-4\pi i j_{ab}/k\}_{a<b} = \times_{a=1}^5 \cD_e\,.
\ee
The remaining restriction is to restrict each of the above $\cD_e$ to the coordinate patch $\cD^r_e$ of $\cM_\Flat(\cS_a,\SU(2))$. Also denote the coordinate patches of $\cM_\Flat(\partial(M_3\backslash \Gamma),\SL(2,\bC))$ and $\cM_\Flat(\partial(M_3\backslash \Gamma),\SU(2))$ as $\cD_\Gamma$ and $\cD^r_\Gamma$ respectively. 
Importantly, one should {\it not} understand the simplicity constraint as restricting $\cD_\Gamma$ to $\cD_\Gamma^r$. The latter space is in fact too small. This is because the FN twist $\{T_{ab}\}$ are not coordinate in $\cD_\Gamma^r$ but in the larger space $\cD_\Gamma$. 

\begin{figure}[h!]
    \centering
    \includegraphics[width=0.3\linewidth]{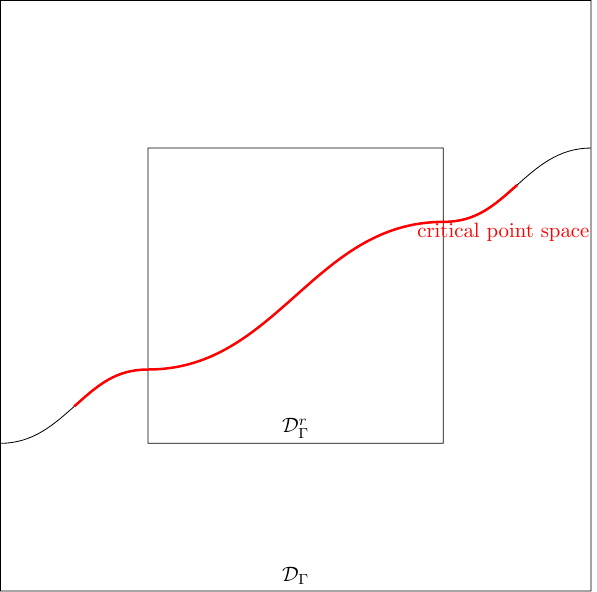}
    \caption{An illustration of the space for the critical points ({\it in red}). It is a subspace of the Lagrangian submanifold $\cB_\Gamma$ ({\it curve in the large box}) but bigger than the Lagrangian submanifold $\cB_\Gamma^r$ ({\it curve in the small box}).}
    \label{fig:Lagrangian}
\end{figure}

To clarify this point in another way, we can now reverse the logic. A point in  $\cD_\Gamma^r$ is characterized by $5V-2E_{\rm in}$ pairs of FG coordinates, denoted as $\{M_i,P_i\}_{i=1}^{5V-2E_{\rm in}}$, and $10V-4E_{\rm in}$ pairs of FN coordinates, denoted as $\{L_{l}=-4\pi i j_{l}/k, T^0_{l}=\in i\R\}_{l=1}^{10V-4E_{\rm in}}$. We embed the phase space $\cD_\Gamma^r$ to $\cD_\Gamma$ by relaxing $T^0_{l}\rightarrow T_l\in \bC$ then restrict to its Lagrange submanifold $\cB_\Gamma$ by imposing the $\SL(2,\bC)$ flatness on the connection. The (complex) dimension of $\cB_\Gamma$ is indeed $\left[2\lb 5V-2E_{\rm in}\rb+2\lb 10V-4E_{\rm in}\rb\right]/2=15V-6E_{\rm in}$. A point in such obtained space corresponds to a critical point of the spinfoam amplitude \eqref{eq:spinfoam_amplitude}. Therefore, the final space for critical points is bigger than $\cB_\Gamma^r$ and is a subspace of $\cB_\Gamma$. This is illustrated in fig.\ref{fig:Lagrangian}.

\section{Conclusion and discussion}
\label{sec:conclusion}

In this paper, we have re-defined the vertex amplitude that is different from the original $\Lambda$-SF model introduced in \cite{Han:2021tzw} by using a new set of canonical symplectic coordinates $\cM_\Flat(\SG,\SL(2,\bC))$ for Chern-Simons theory on $\SG$. 
It requires relaxing the symplectic transformation to be within ${\rm Sp}(30,\f{\Z}{4})$, and the Chern-Simons partition function is well-defined when the Chern-Simons level $k$ is 8 times positive integer, \eg $k\in 8\N$. 
The new set of the symplectic coordinates has a simple geometrical interpretation, which in turn makes it possible to define a universal face amplitude as a simple generalization of that in the EPRL model.  
We have also pushed the analysis to 4-complexes whose spinfoam graphs are colorable that are used in the colored tensor models. 

A direct generalization is to consider the spinfoam amplitude for any 4-complexes not restricted to those with colorable spinfoam graphs. To this end, one should define new edge amplitudes that presumably take more involved forms. Given two convex tetrahedra, say $\tetra_a$ and $\tetra_b$ attached with FN and FG coordinates $\{\{L_{ac}\}_c, M_a,P_a\}$ and  $\{\{L_{bd}\}_d, M_b,P_b\}$ respectively, there are 12 ways to glue the boundary triangles pairwise, each of which gives a set of 6 gluing constraints as the linear equations of these coordinates. These gluing constraints are reflected in the definition of edge amplitude. We leave this for future investigation. 

By performing the stationary phase analysis, we have found the complete set of critical points, which give the geometry of constantly flat 4-complexes as in the original $\Lambda$-SF model. 
The geometrical interpretation of the critical points implies that one can directly compute the semi-classical approximation of the spinfoam amplitude given the geometrical data set and the boundary condition without going through the stationary phase analysis. This is investigated in \cite{Pan:2025sut}. 

The need to consider Chern-Simons theory at level $k\in 8\N$ seems to be related to the complexity of the graph as $k$ can be a general integer in the case of Chern-Simons theory on a knot-complement 3-manifolds \cite{Dimofte:2011gm,Dimofte:2011ju}. It would be interesting to study the Chern-Simons theory with the same techniques in this paper on other graph-complement 3-manifolds and investigate how the graph imposes restrictions on the value of $k$. This could bring insights to the broader field of quantum Chern-Simons theory.  

Moreover, since the spinfoam amplitude defined in this work is valid for any 4-complex consistent with the colored tensor model, it is also worthwhile to investigate the CGFT formulation of the spinfoam model using this construction. A CGFT-based approach would provide a field-theoretic reformulation of the spinfoam model with $\Lambda\neq 0$, relaxing the dependence on a fixed spacetime triangulation. Ultimately, this could enable the application of statistical field theory methods and renormalization group techniques to loop quantum gravity, potentially leading to new insights into the renormalization properties of spinfoam models.

\begin{acknowledgements}
This work receives support from the National Science Foundation through grants PHY-2207763, the Blaumann Foundation, the College of Science Research Fellowship at Florida Atlantic University and the Jumpstart Postdoctoral Program at Florida Atlantic University.
\end{acknowledgements}

\appendix
\renewcommand\thesection{\Alph{section}}

\section{Symplectic transformation of coordinates on $\SG$}
\label{app:symplec_tranf}

In this appendix, we give the explicit expressions for the sub-matrices $\bA,\bB,{\bf C},{\bf D}$ of the symplectic matrix $\bM$ in \eqref{eq:symp_transf} and \eqref{eq:symplectic_decompse} and vectors $\vec{t}_\alpha$, $\vec{t}_\beta$. We also give detailed expressions on the FN coordinates in terms of the coordinates $(\vec{\Phi},\vec{\Pi})$ on the ideal octahedra before the symplectic transformations. 

\begin{subequations}
\begin{align}
\bA=&\left(
\begin{array}{ccccccccccccccc}
 0 & 0 & 0 & 0 & 0 & 0 & 0 & -1 & -1 & 0 & -1 & -1 & 0 & -1 & -1 \\
 0 & 0 & 0 & 0 & -1 & -1 & 0 & 0 & 0 & 0 & -1 & 1 & 0 & 0 & 0 \\
 0 & 0 & 0 & 0 & -1 & 1 & 0 & 0 & 0 & 0 & 0 & 0 & 0 & -1 & 1 \\
 0 & 0 & 0 & 0 & 0 & 0 & 0 & -1 & 1 & 0 & 0 & 0 & 0 & 0 & 0 \\
 2 & 1 & 1 & 0 & 0 & 0 & 0 & 0 & 0 & 0 & 0 & 0 & 2 & 1 & 1 \\
 0 & 1 & 1 & 0 & 0 & 0 & 2 & 1 & 1 & 0 & 0 & 0 & 0 & 0 & 0 \\
 0 & 0 & 0 & 0 & 0 & 0 & 0 & 0 & 0 & 2 & 1 & 1 & 0 & 0 & 0 \\
 0 & 0 & 0 & 0 & 0 & 0 & 0 & 0 & 0 & 0 & 0 & 0 & 0 & 1 & 1 \\
 0 & 1 & 1 & 2 & 1 & 1 & 0 & 0 & 0 & 0 & -1 & -1 & 0 & 0 & 0 \\
 0 & -1 & 1 & 0 & -1 & -1 & 0 & 1 & 1 & 0 & 0 & 0 & 0 & 0 & 0 \\
 0 & 0 & 0 & 0 & -\frac{1}{2} & -\frac{1}{2} & 0 & \frac{1}{2} & -\frac{1}{2} & 0 & 0 & 1 & 0 & -\frac{1}{2} & \frac{1}{2} \\
 0 & \frac{1}{2} & \frac{1}{2} & 0 & 0 & 0 & 0 & 0 & 0 & 0 & \frac{1}{2} & \frac{1}{2} & 1 & \frac{1}{2} & \frac{1}{2} \\
 -1 & -1 & -1 & 0 & \frac{1}{2} & \frac{1}{2} & 0 & 0 & 0 & 0 & \frac{1}{2} & -\frac{1}{2} & -1 & -\frac{1}{2} & -\frac{1}{2} \\
 0 & 0 & 1 & 0 & -\frac{1}{2} & -\frac{1}{2} & 0 & 0 & 0 & 0 & 0 & 0 & 0 & 1 & 0 \\
 0 & -\frac{1}{2} & -\frac{1}{2} & 0 & \frac{1}{2} & \frac{1}{2} & 0 & \frac{1}{2} & -\frac{1}{2} & -1 & -\frac{1}{2} & -\frac{1}{2} & 0 & 0 & 0 \\
\end{array}
\right)\,,
\label{eq:A}\\
\bB=&\left(
\begin{array}{ccccccccccccccc}
 0 & 0 & 0 & 0 & 0 & 0 & 0 & -1 & 0 & 0 & -1 & 0 & 0 & -1 & 0 \\
 0 & 0 & 0 & 0 & -1 & 0 & 0 & 0 & 0 & 0 & -1 & 1 & 0 & 0 & -1 \\
 0 & 0 & 0 & 0 & -1 & 1 & 0 & 0 & -1 & 0 & 0 & 0 & 0 & -1 & 1 \\
 0 & 0 & 0 & 0 & 0 & -1 & 0 & -1 & 1 & 0 & 0 & -1 & 0 & 0 & 0 \\
 1 & 0 & 0 & 0 & 0 & 0 & 0 & 0 & 0 & -1 & 1 & 0 & 1 & 0 & 0 \\
 0 & 1 & 0 & 0 & 0 & 0 & 1 & 0 & 0 & 0 & 0 & 0 & -1 & 1 & 0 \\
 -1 & 1 & 0 & 0 & 0 & 0 & -1 & 1 & 0 & 1 & 0 & 0 & 0 & 0 & 0 \\
 0 & 0 & 1 & -1 & 1 & 0 & 0 & 0 & 0 & 0 & 0 & 0 & -1 & 0 & 1 \\
 -1 & 0 & 1 & 1 & 0 & 0 & 0 & 0 & 0 & 1 & 0 & -1 & 0 & 0 & 0 \\
 0 & -1 & 1 & 1 & 0 & -1 & -1 & 0 & 1 & 0 & 0 & 0 & 0 & 0 & 0 \\
 0 & 0 & 0 & 0 & -\frac{1}{2} & 0 & 0 & \frac{1}{2} & -1 & 0 & 0 & \frac{1}{2} & 0 & -\frac{1}{2} & \frac{1}{2} \\
 0 & \frac{1}{2} & 0 & 0 & 0 & 0 & -\frac{1}{2} & \frac{1}{2} & 0 & 0 & \frac{1}{2} & 0 & \frac{1}{2} & 0 & 0 \\
 0 & 0 & -\frac{1}{2} & \frac{1}{2} & 0 & 0 & 0 & 0 & 0 & 0 & \frac{1}{2} & -\frac{1}{2} & -\frac{1}{2} & 0 & 0 \\
 0 & 0 & \frac{1}{2} & \frac{1}{2} & 0 & -\frac{1}{2} & 0 & 0 & \frac{1}{2} & 0 & 0 & 0 & -\frac{1}{2} & \frac{1}{2} & 0 \\
 \frac{1}{2} & 0 & -\frac{1}{2} & -\frac{1}{2} & 0 & \frac{1}{2} & 0 & \frac{1}{2} & -\frac{1}{2} & -\frac{1}{2} & 0 & 0 & 0 & 0 & 0 \\
\end{array}
\right)\,,
\label{eq:B}\\
{\bf C}=&\left(
\begin{array}{ccccccccccccccc}
 0 & 0 & 0 & 0 & 0 & 0 & \frac{1}{2} & \frac{1}{2} & 0 & 0 & 0 & 0 & 0 & 0 & 0 \\
 0 & 0 & 0 & 0 & 0 & 0 & 0 & 0 & 0 & 0 & 0 & 0 & \frac{1}{2} & \frac{1}{2} & 1 \\
 0 & 0 & 0 & 0 & 0 & 0 & 0 & 0 & 0 & 0 & 0 & 0 & 0 & 0 & 0 \\
 0 & 0 & 0 & 0 & 0 & 0 & 0 & 0 & 0 & \frac{1}{2} & \frac{1}{2} & 1 & 0 & 0 & 0 \\
 0 & 0 & 0 & 0 & 0 & 0 & 0 & 0 & 0 & \frac{1}{2} & -\frac{1}{2} & 0 & 0 & 0 & 0 \\
 0 & 0 & 0 & 0 & 0 & 0 & 0 & 0 & 0 & 0 & 0 & 0 & \frac{1}{2} & -\frac{1}{2} & 0 \\
 \frac{1}{2} & -\frac{1}{2} & 0 & 0 & 0 & 0 & 0 & 0 & 0 & 0 & 0 & 0 & 0 & 0 & 0 \\
 -\frac{1}{2} & -\frac{1}{2} & -1 & 0 & 0 & 0 & 0 & 0 & 0 & 0 & 0 & 0 & 0 & 0 & 0 \\
 0 & 0 & 0 & 0 & 0 & 0 & 0 & 0 & 0 & 0 & 0 & 0 & 0 & 0 & 0 \\
 0 & 0 & 0 & 0 & 0 & 0 & \frac{1}{2} & \frac{1}{2} & 0 & 0 & 0 & 0 & 0 & 0 & 0 \\
 0 & 0 & 0 & 0 & \frac{1}{2} & \frac{1}{2} & 0 & -\frac{1}{2} & \frac{1}{2} & 0 & -\frac{1}{2} & -\frac{1}{2} & 0 & 0 & -1 \\
 -1 & -\frac{1}{2} & -\frac{1}{2} & 0 & 0 & 0 & 1 & \frac{1}{2} & \frac{1}{2} & -1 & -\frac{1}{2} & -\frac{1}{2} & -1 & 0 & 0 \\
 -1 & -\frac{1}{2} & -\frac{1}{2} & 0 & \frac{1}{2} & \frac{1}{2} & 0 & 0 & 0 & 0 & -\frac{1}{2} & -\frac{1}{2} & 0 & \frac{1}{2} & \frac{1}{2} \\
 0 & -\frac{1}{2} & -\frac{1}{2} & 0 & \frac{1}{2} & \frac{1}{2} & -1 & -\frac{1}{2} & -\frac{1}{2} & 0 & 0 & 0 & 0 & -\frac{1}{2} & -\frac{1}{2} \\
 0 & -1 & 0 & 0 & 0 & 0 & 0 & \frac{1}{2} & -\frac{1}{2} & 0 & \frac{1}{2} & \frac{1}{2} & 0 & 0 & 0 \\
\end{array}
\right)
\label{eq:C}\\
{\bf D}=&\left(
\begin{array}{ccccccccccccccc}
 0 & 0 & 0 & 0 & 0 & 0 & \frac{1}{2} & 0 & -\frac{1}{2} & 0 & 0 & 0 & 0 & 0 & 0 \\
 0 & 0 & 0 & 0 & 0 & 0 & 0 & 0 & 0 & 0 & 0 & 0 & 0 & 0 & \frac{1}{2} \\
 0 & 0 & 0 & 0 & -\frac{1}{2} & \frac{1}{2} & 0 & 0 & 0 & 0 & 0 & 0 & 0 & 0 & 0 \\
 0 & 0 & 0 & 0 & 0 & 0 & 0 & 0 & 0 & 0 & 0 & \frac{1}{2} & 0 & 0 & 0 \\
 0 & 0 & 0 & 0 & 0 & 0 & 0 & 0 & 0 & \frac{1}{2} & -\frac{1}{2} & 0 & 0 & 0 & 0 \\
 0 & 0 & 0 & 0 & 0 & 0 & 0 & 0 & 0 & 0 & 0 & 0 & \frac{1}{2} & -\frac{1}{2} & 0 \\
 \frac{1}{2} & -\frac{1}{2} & 0 & 0 & 0 & 0 & 0 & 0 & 0 & 0 & 0 & 0 & 0 & 0 & 0 \\
 0 & 0 & -\frac{1}{2} & 0 & 0 & 0 & 0 & 0 & 0 & 0 & 0 & 0 & 0 & 0 & 0 \\
 0 & 0 & 0 & \frac{1}{2} & 0 & 0 & 0 & 0 & 0 & 0 & 0 & 0 & 0 & 0 & 0 \\
 0 & 0 & 0 & 0 & 0 & 0 & 0 & \frac{1}{2} & 0 & 0 & 0 & 0 & 0 & 0 & 0 \\
 0 & 0 & 0 & 0 & \frac{1}{2} & -\frac{1}{2} & 0 & -\frac{1}{2} & \frac{1}{2} & 0 & -\frac{1}{2} & 0 & 0 & 0 & -\frac{1}{2} \\
 -\frac{1}{2} & 0 & 0 & 0 & 0 & 0 & 1 & -\frac{1}{2} & 0 & -\frac{1}{2} & 0 & 0 & -\frac{1}{2} & \frac{1}{2} & 0 \\
 -\frac{1}{2} & 0 & 0 & 0 & \frac{1}{2} & 0 & 0 & 0 & 0 & \frac{1}{2} & 0 & -\frac{1}{2} & -\frac{1}{2} & 0 & \frac{1}{2} \\
 0 & -\frac{1}{2} & 0 & 0 & -\frac{1}{2} & \frac{1}{2} & -\frac{1}{2} & 0 & -\frac{1}{2} & 0 & 0 & 0 & \frac{1}{2} & 0 & -\frac{1}{2} \\
 \frac{1}{2} & -\frac{1}{2} & 0 & 0 & 0 & \frac{1}{2} & -\frac{1}{2} & 1 & -\frac{1}{2} & -\frac{1}{2} & 0 & \frac{1}{2} & 0 & 0 & 0 \\
\end{array}
\right)
\end{align}
\label{eq:AB}
\end{subequations}
The vectors $\vec{t}_\alpha$ and $\vec{t}_\beta$ for affine translations are
\be
\vec{t}_\alpha=\lb3,1,0,0,-4,-3,-2,-1,-2,0,1,0,3,\frac{1}{2},3\rb^\top\,,\quad 
\vec{t}_\beta=\lb-\frac{1}{2},-1,0,-1,0,0,0,1,0,-\frac{1}{2},\frac{3}{2},\frac{1}{2},\frac{5}{2},\frac{5}{2},1\rb^\top\,.
\label{eq:t_alpha_t_beta}
\ee

The FG coordinates $\{\chi_{bc}^{(a)}\}_{a,b,c}$ and coordinates $\{\{L_{ab}\}_{b},M_a,P_a\}$ are related by linear combination with integer prefactors as follows. 
\be
\ba{lll}
\chi^{(1)}_{23}=-L_{12}+L_{14}+P_1\,,\quad
&\chi^{(1)}_{24}=-L_{14}+L_{15}+M_1\,,\quad
&\chi^{(1)}_{25}=L_{13}+L_{14}-M_1-P_1+3 i \pi \,,\\[0.15cm]
\chi^{(1)}_{34}=L_{12}+L_{15}-M_1-P_1+3 i \pi \,,\quad
&\chi^{(1)}_{35}=L_{12}-L_{13}+M_1\,,\quad
&\chi^{(1)}_{45}=L_{13}-L_{15}+P_1\,,\\[0.35cm]
\chi^{(2)}_{13}=L_{12}+L_{25}+M_2+P_2\,,\quad
&\chi^{(2)}_{14}=L_{23}+L_{25}-M_2+3 i \pi\,,\quad
&\chi^{(2)}_{15}=L_{24}-L_{25}-P_2\,,\\[0.15cm]
\chi^{(2)}_{34}=-L_{12}-L_{23}-P_2\,,\quad
&\chi^{(2)}_{35}=-L_{12}+L_{24}-M_2+3 i \pi\,,\quad
&\chi^{(2)}_{45}=L_{23}-L_{24}+M_2+P_2\,,\\[0.35cm]
\chi^{(3)}_{12}=L_{34}+L_{35}-P_3+3 i \pi\,,\quad
&\chi^{(3)}_{14}=-L_{23}-L_{34}-M_3+P_3\,,\quad
&\chi^{(3)}_{15}=L_{13}+L_{34}+M_3\,,\\[0.15cm]
\chi^{(3)}_{24}=L_{23}+L_{35}+M_3\,,\quad
&\chi^{(3)}_{25}=-L_{13}-L_{35}-M_3+P_3\,,\quad
&\chi^{(3)}_{45}=-L_{13}-L_{23}-P_3+3 i \pi\,,\\[0.35cm]
\chi^{(4)}_{12}=L_{14}-L_{34}+M_4\,,\quad
&\chi^{(4)}_{13}=L_{34}+L_{45}+P_4\,,\quad
&\chi^{(4)}_{15}=-L_{24}-L_{34}-M_4-P_4+3 i \pi \,,\\[0.15cm]
\chi^{(4)}_{23}=-L_{14}+L_{45}-M_4-P_4+3 i \pi\,,\quad
&\chi^{(4)}_{25}=-L_{14}+L_{24}+P_4\,,\quad
&\chi^{(4)}_{35}=-L_{24}-L_{45}+M_4\,,\\[0.35cm]
\chi^{(5)}_{12}=L_{25}-L_{35}+M_5-P_5\,,\quad
&\chi^{(5)}_{13}=-L_{25}-L_{45}-M_5+3 i \pi\,,\quad
&\chi^{(5)}_{14}=L_{15}-L_{25}+P_5\,,\\[0.15cm]
\chi^{(5)}_{23}=L_{35}-L_{45}+P_5\,,\quad
&\chi^{(5)}_{24}=-L_{15}-L_{35}-M_5+3 i \pi\,,\quad
&\chi^{(5)}_{34}=-L_{15}+L_{45}+M_5-P_5\,.
\ea
\label{eq:chi_to_LMP}
\ee
The FN lengths $2L_{ab}$ and FN twists $T_{ab}$ are calculated using the snake rule on cusps \cite{Dimofte:2011gm} \footnote{A good reference for computing $T_{ab}$ to get the results \eqref{eq:T} is Appendix H of \cite{Han:2023hbe}, noting that the role of $\fz'_i$ and $\fz''_i$ ($\fz_i=x_i,y_i,z_i,w_i$) are exchanged therein compared to this paper.}.
Explicitly,
\be
\begin{aligned}
&  2L_{12}=\chi^{(1)}_{34}+\chi^{(1)}_{35}+\chi^{(1)}_{45}-3i\pi=-P_{Y_3}-P_{Y_4}-P_{Y_5}-Y_3-Y_4-Y_5-Z_3-Z_4-Z_5+3 i \pi \,,\\
&  2L_{13}=\chi^{(1)}_{24}+\chi^{(1)}_{25}+\chi^{(1)}_{45}-3i\pi=-P_{Y_2}-P_{Y_4}+P_{Z_4}-P_{Z_5}-Y_2-Y_4-Z_2+Z_4+i \pi\,, \\
&  2L_{14}=\chi^{(1)}_{23}+\chi^{(1)}_{25}+\chi^{(1)}_{35}-3i\pi=-P_{Y_2}-P_{Y_5}+P_{Z_2}-P_{Z_3}+P_{Z_5}-Y_2-Y_5+Z_2+Z_5\,, \\
&  2L_{15}=\chi^{(1)}_{23}+\chi^{(1)}_{24}+\chi^{(1)}_{34}-3i\pi=-P_{Y_3}-P_{Z_2}+P_{Z_3}-P_{Z_4}-Y_3+Z_3\,, \\
&  2L_{23}=\chi^{(2)}_{14}+\chi^{(2)}_{15}+\chi^{(2)}_{45}-3i\pi=P_{X_1}-P_{X_4}+P_{X_5}+P_{Y_4}+2 X_1+2 X_5+Y_1+Y_5+Z_1+Z_5-4 i \pi\,, \\
&  2L_{24}=\chi^{(2)}_{13}+\chi^{(2)}_{15}+\chi^{(2)}_{35}-3i\pi=P_{X_3}-P_{X_5}+P_{Y_1}+P_{Y_5}+2 X_3+Y_1+Y_3+Z_1+Z_3-3 i \pi\,, \\
&  2L_{25}=\chi^{(2)}_{13}+\chi^{(2)}_{14}+\chi^{(2)}_{34}-3i\pi=-P_{X_1}-P_{X_3}+P_{X_4}+P_{Y_1}+P_{Y_3}+2 X_4+Y_4+Z_4-2 i \pi \,, \\
&  2L_{34}=\chi^{(3)}_{12}+\chi^{(3)}_{15}+\chi^{(3)}_{25}-3i\pi=-P_{X_2}-P_{X_5}+P_{Y_2}+P_{Z_1}+P_{Z_5}+Y_5+Z_5-i \pi \,, \\
&  2L_{35}=\chi^{(3)}_{12}+\chi^{(3)}_{14}+\chi^{(3)}_{24}-3i\pi=-P_{X_1}+P_{X_2}+P_{X_4}+P_{Z_1}-P_{Z_4}+2 X_2+Y_1+Y_2-Y_4+Z_1+Z_2-Z_4-2 i \pi\,, \\
&  2L_{45}=\chi^{(4)}_{12}+\chi^{(4)}_{13}+\chi^{(4)}_{23}-3i\pi=P_{X_2}-P_{X_3}-P_{Y_1}+P_{Z_1}-P_{Z_2}+P_{Z_3}-Y_1-Y_2+Y_3+Z_1-Z_2+Z_3\,.
\end{aligned}
\label{eq:L}
\ee
\be
\begin{aligned}
&  T_{12}=\frac{1}{2} \left(-X'_3+Y_3-Z''_3\right)=\frac{1}{2} \left(P_{X_3}-P_{Z_3}+X_3+Y_3-i \pi \right)\,\,
&&  T_{13}=\frac{1}{2} \left(W_5-Z'_5\right)=\frac{1}{2} \left(P_{Z_5}+X_5+Y_5+2 Z_5-2 i \pi \right)\,, \\
&  T_{14}=\frac{1}{2} \left(Z''_2-Y''_2\right)=\frac{1}{2} \left(P_{Z_2}-P_{Y_2}\right) \,,\,
&&  T_{15}=\frac{1}{2} \left(W_4-Z'_4\right)=\frac{1}{2} \left(P_{Z_4}+X_4+Y_4+2 Z_4-2 i \pi \right)\,, \\
&  T_{23}=\frac{1}{2} \left(Y'_4-X'_4\right)=\frac{1}{2} \left(P_{X_4}-P_{Y_4}+X_4-Y_4\right)\,, \,
&&  T_{24}=\frac{1}{2} \left(Y'_5-X'_5\right)=\frac{1}{2} \left(P_{X_5}-P_{Y_5}+X_5-Y_5\right)\,, \\
&  T_{25}=\frac{1}{2} \left(Y'_1-X'_1\right)=\frac{1}{2} \left(P_{X_1}-P_{Y_1}+X_1-Y_1\right) \,, \,
&&  T_{34}=\frac{1}{2} \left(Z'_1-W_1\right)=\frac{1}{2} \left(-P_{Z_1}-X_1-Y_1-2 Z_1+2 i \pi \right)\,, \\
&  T_{35}=\frac{1}{2} \left(X''_2-W''_2\right)=\f12P_{X_2}\,, \,
&&  T_{45}=\frac{1}{2} \left(W_3+Y''_3-Z_3\right)=\frac{1}{2} \left(P_{Y_3}+X_3+Y_3-i \pi \right)\,.
\end{aligned}
\label{eq:T}
\ee

\bibliographystyle{bib-style} 
\bibliography{SFC.bib}

\end{document}